\documentclass[a4paper,11pt]{article}
\pdfoutput=1 

\usepackage{jcappub} 

\usepackage[T1]{fontenc} 
\usepackage{multirow}

\begin{document}

\title{{Influence of pressure anisotropy on mass-radius relation and stability of millisecond pulsars in $f(Q)$ gravity}}                      

\author[a]{S. K. Maurya,}
\author[b]{Ksh. Newton Singh,}
\author[c]{G. Mustafa,}
\author[d,e]{M. Govender,}
\author[f]{Abdelghani Errehymy,}
\author[a]{Abdul Aziz}

\affiliation[a]{Department of Mathematics and Physical Sciences,  University of Nizwa, Nizwa 616, Oman}
\affiliation[b]{Department of Physics, National Defence Academy, Khadakwasla, Pune 411023, India}
\affiliation[c]{Department of Physics,
Zhejiang Normal University, Jinhua 321004, People’s Republic of China}
\affiliation[d]{Department of Mathematics, Durban University of Technology, Durban 4000, South Africa}
\affiliation[e]{Institute of Systems Science, Durban University of Technology, Durban 4000, South Africa}
\affiliation[f]{Astrophysics Research Centre, School of Mathematics, Statistics and Computer Science, University of KwaZulu-Natal, Private Bag X54001, Durban 4000, South Africa}
\emailAdd{sunil@unizwa.edu.om}
\emailAdd{ntnphy@gmail.com}
\emailAdd{gmustafa3828@gmail.com}
\emailAdd{megandhreng@dut.ac.za}
\emailAdd{abdelghani.errehymy@gmail.com}
\emailAdd{azizmail2012@gmail.com}

\date{\today}
\abstract{In this study we explore the astrophysical implications of pressure anisotropy on the physical characteristics of millisecond pulsars within the framework of $f(Q)$ gravity, {in particular $f(Q)=-\alpha\, Q - \beta$, where $\alpha$ and $\beta$ are constants.} Starting off with the field equations for anisotropic matter configurations, we adopt the physically salient Durgapal-Fuloria ansatz together with a well-motivated anisotropic factor for the interior matter distribution. This leads to a nonlinear second order differential equation which is integrated to give the complete gravitational and thermodynamical properties of the stellar object. The resulting model is subjected to rigorous  tests to ensure that it qualifies as a physically viable compact object within the $f(Q)$-gravity framework. We study in detail the impact of anisotropy on the mass, radius and stability of the star. Our analyses indicate that our models are well-behaved, singularity-free and can account for the existence of a wide range of observed pulsars with masses ranging from 2.08 to 2.67 $M_{\odot}$, with the upper value being in the so-called {\em mass gap} regime observed in gravitational events such as GW190814. {A comparison of the so-called {\em Symmetric Teleparallel Equivalent to GR} (STEGR) models with classical General Relativity (GR) models reveal that the anisotropy parameter and the sign of $\beta$ impact on the predicted radii of pulsars. In particular, STEGR models have larger radii than their GR counterparts.} } 


\maketitle
\flushbottom

\section{Introduction}

The first detection of a pulsar, aptly named LGM-1, by Jocelyn Bell in 1967 created quite a stir amongst her collaborators \citep{bell}. The regularity of the signals emanating from the source led them to believe that it could be artificially generated, hence the reference to Little Green Men 1 or LGM-1. Today we know that Bell had discovered a pulsar which is a rapidly rotating neutron star with a period of 1.3373 seconds. The neutron star itself is born out of a cataclysmic supernova explosion. Pulsars occupy an important place in observational astrophysics as they serve as natural laboratories for probing the nature of superdense matter, gravitational waves~\citep{waves1,waves2,waves3} and the end-states of gravitational collapse. Pulsars such as Cen~X-$3$, SAX~J$1748.9-2021$, Vela~X-$1$ and PSR~J$0030+0451$ have been extensively studied in the literature. Cen-X-$3$ was serendipituously discovered when rocket-borne detectors were trained upon X-ray sources Sco XR-$1$ and Tau XR-$1$. X-rays were detected from the direction
of the constellations Vela and Lupus in which the sources 
Vel XR-$1$ and Lup XR-$1$ were respectively located with the new source Cen-X-$3$ inhabiting the constellation Centaurus~\citep{x3}.

The pioneering work of Bowers \& Liang \cite{BL1974} cast light on the impact of anisotropic stresses in relativistic compact objects. This seminal paper extended Bondi's earlier work on isotropic configurations to include pressure anisotropy without the imposition of an equation of state (EOS). The key findings of their investigation were the prediction of higher surface redshifts and the maximum allowable mass of gravitating bodies in the presence of unequal radial and transverse stresses. It was well-known that physical processes such as pion condensation \citep{saw1,saw2}, neutrino transport \citep{mart1}, superconducting states, amongst others, give rise to anisotropy within the stellar core. These processes play a significant role at ultra-high densities, especially during the last stages of gravitational collapse. In their review article, Herrera and Santos highlighted the impact of anisotropy in radiating stars~\citep{local}. The highlight of their work signified the modification of the adiabatic index in both the Newtonian and post-Newtonian approximations which generalises the classic stability result obtained by Chandrasekhar for isotropic matter distributions~\citep{chandra}. The dynamical (in)stability of the collapsing body is affected by the sign of the anisotropic factor as well as the gradient of the radial pressure. In a more recent study, Herrera demonstrated the instability of the pressure isotropy condition in fluids losing hydrostatic equilibrium~\citep{pi}. It was shown that an initial isotropic matter distribution, upon leaving hydrostatic equilibrium, will evolve into an anisotropic regime. In particular, the appearance of the radial pressure term in the TOV equation (which is absent in the Newtonian analogue) plays a crucial role in the anisotropisation of the collapsing fluid.  

Recently, there has been an exponential growth of exact solutions of the Einstein field equations describing anisotropic compact objects.  A popular approach to close the system of coupled equations is to employ a metric ansatz for one of the gravitational potentials and impose a governing equation that dictates the symmetry and/or nature of the spacetime (for example, the Karmarkar condition for embedding class I spacetimes~\citep{k1,k2,k3,k4}), an EOS (which relates the radial pressure and energy density), complexity-free condition, amongst others. Various EOS's ranging from the simple linear EOS ($p_r = \psi \rho$), through to the MIT bag model, the polytropic EOS and the CFL EOS have been employed to complete the gravitational and thermodynamical behaviour of the models. The sophistication of the plethora of EOS's has grown with experimental results from particle physics. Numerous stellar models arising from the imposition of an EOS have revealed the impact of the bag constant, quark interactions and quark energies. In a recent article, the authors considered quarks to form Cooper pairs which obeyed a color–flavor-locked (CFL) EOS which formed the stellar fluid of strange stars admitting masses in the vicinity of 3.61 $M_{\odot}$~\citep{cfl1}.  They further provided evidence that the 3-flavor strange quark matter (SQM) in the CFL phase exhibits complete stability in comparison to $^{56}Fe$ for strange quark masses, $m_s < 228.3 MeV$. In addition, the stability of SQM increases with a decrease in $m_s$. Their study of observed compact objects showed that PSR J1614-2230 and PSR J0740+6620 are found to be anisotropic in nature. In a more recent study, Mohanty et al. employed a total of 60 EOS's to investigate the impact of anisotropy in neutron stars~\citep{Mohanty}. By using a numerical methodology, the researchers demonstrated the feasibility of producing neutron star masses above 2.0 $M_{\odot}$ within the framework of GR by influencing the magnitude of anisotropy present in the stellar core. 

The inclusion of pressure anisotropy via gravitational decoupling (GD) has been successfully implemented to produce a wide spectrum of compact stellar objects. The basis of the GD framework is to introduce an additional source via the energy-momentum tensor of the standard Einstein field equations. The additional source term mimics anisotropy within the stellar configuration. The so-called minimal geometric deformation (MGD) technique~\citep{o1} and its generalisation, referred to as complete geometric deformation (CGD) \citep{o2} have led to anisotropic analogues of well-known isotropic solutions of the Einstein field equations. It must be pointed that while the isotropic seed solution may suffer various pathologies, their anisotropic counterparts obtained via MGD or CGD may represent realistic compact stellar objects. It has been shown that the decoupling constant influences stellar characteristics such as mass-radius relations, stability and the upper mass limit of compact objects~\citep{gove1,gove2,gove3}. 

Inspired by the observations of the LIGO-VIRGO collaboration of events such as GW170817 and GW190814, researchers have renewed their efforts in modeling the compact objects involved in binary mergers which are sources for gravitational waves.  In particular, the gravitational wave event GW190814 indicates that the signals were generated by a fusion of a black hole with a mass ranging from 22.2 to 24.3$ M_{\odot}$ and a compact object with a mass ranging from 2.50 to 2.67$M_{\odot}$. On the other hand, the GW170817 event is ascribed to the merger of two neutron stars with masses in the range 0.86 - 2.26 $M_{\odot}$. In order to achieve stellar masses greater than 2 $M_{\odot}$ in standard GR without invoking exotic matter distributions or rotation, theorists have ventured into modified gravity theories. One such popular theory is the $f(Q)$-gravity which has proved to be quite diverse in its predictive power, both in the astrophysical and cosmological frontiers. {A seminal study of self-gravitating objects in $f(Q)$ gravity is due to Wang et al. \cite{wanga}. In this work they studied spherically symmetric fluid spheres with anisotropic pressure within the stellar interior. They demonstrated the existence of the Schwarzschild (anti-)de Sitter solution and further showed that there is no exact Schwarzschild solution for nontrivial $f(Q)$ functions.} A power-law form for $f(Q) = \alpha + \beta \,Q^n$ was investigated by Capozziello and D'Agostino \citep{Capoplb}. A special case of the Capozziello and D'Agostino ansatz, i.e., a linear form of $f(Q)$ (taking $n = 1$) was employed by Maurya et al. \cite{mauryajcap} to study compact objects with anisotropic pressure by the anisotropisation of the Tolman IV solution via gravitational decoupling. The ensuing models described physically viable stellar structures including a range of stars endowed with masses of the order of 1.2 - 2.26 $M_{\odot}$. In particular, this work demonstrated that contributions from the nonmetricity factor predict larger stars. By adopting a singularity-free form for the gravitational potentials described by the Tolman-Kuchowicz ansatz, Bhar et al. modeled anisotropic hybrid stars in which the stellar fluid comprised of a superposition of strange quark matter (SQM) and ordinary baryonic matter (OBM) distributions within the $f(Q)$ gravity framework \citep{Bhar2023}. In addition, they adopted the MIT bag model EOS to complete the gravitational behaviour of the model. Their models were singularity-free and covered a range of stellar masses including compact objects in the range required by the secondary component of the GW 190814 event~\citep{Bhar11}. The mass-gap conundrum arising in gravitational events has presented researchers with various challenges over the recent years. The latest candidate is a pulsar of mass 2.09 to 2.71 $M_{\odot}$ which forms one component of a binary system observed by the MeerKat survey \citep{meer}. 

In a recent paper, Maurya et al. \cite{mauryaaps} employed a quadratic EOS together with a linear form for $f(Q)$ to study compact objects within the MGD framework. They showed that the most robust model which accounts for a wide spectrum of stars, especially compact objects with masses in the regime predicted by GW190814 is the one incorporating a superposition of linear and quadratic contributions from the stellar density. In order to study the influence of the quadratic term arising in $f(Q) = Q + aQ^2$, where in the limiting case of $a = 0$, we regain classical GR, Bhar et al. modeled charged compact objects with anisotropic stresses in the interior~\citep{Bhar2024}. They demonstrated that their models were sensitive to the EOS parameter, metricity, charge, and the bag constant. In addition, they noted that the quadratic contribution from $f(Q)$ results in the lowering of the density, radial pressure, electric field, and sound speeds. The novelty of their work lies in the ability of increasing the magnitude of the anisotropy via the quadratic presence in $f(Q)$ without invoking gravitational decoupling or any exotic scalar field or matter distributions, including dark energy. For a recent review on $f(Q)$ gravity, we refer the reader to \cite{Heisenberg2024} for more details. In their recent work, Capozziello et al. \cite{Capozziello2024} have demonstrated that gravitational waves generated in non-metricity-based $f(Q)$ gravity exhibit similar characteristics to those in torsion-based $f(T)$ gravity \cite{Capozziello:2015rda,Bamba:2014zra,Nashed:2011fz,Iorio:2012cm,Awad:2017sau}. Consequently, distinguishing between these gravitational waves and those predicted by General Relativity based solely on wave polarization measurements becomes challenging. This distinction contrasts with the behavior observed in curvature-based $f(R)$ gravity \cite{Nojiri:2006ri,Sotiriou:2008rp,DeFelice:2010aj,Nojiri:2010wj,Nojiri:2017ncd,Astashenok:2020qds,Astashenok:2021peo,Malik1}, where an additional scalar mode is always present when $f(R) \neq R$. In addition, some pioneering works done by Ray and his collaborators can be seen in following references~\cite{Ray1,Ray2,Ray3,Ray4}. Furthermore, Capozziello et al. also performed some work on metric affine gravity~\cite{Capozziello11,Capozziello12}. {The evolution of cosmological models in $f(Q)$ gravity and modified gravity theories can be found in excellent review articles and foundational studies in these works \cite{H1,H2,H3}. In order to elucidate the number of physical degrees of freedom arising in $f(Q)$ gravity and their dependence on the functional form of $f$ various studies which  adopted a Hamiltonian analysis based on the Dirac-Bergmann algorithm can be found in \cite{deg1,deg2,deg3}. A beautiful review of $f(Q)$ gravity threading aspects of cosmology, black holes and degrees of freedom can be found in \cite{H4}. The consequences of the physical degrees of freedom within the $f(Q)$ framework and their link to ghost scalar modes have been investigated by Hu et al. \cite{K1}. }

{ It is thought that dS spacetime  or a non-zero  positive cosmological constant ($\Lambda$) causes the accelerated expansion of the universe which is challenging to incorporate  in theories of supergravity and superstrings.  The quantum physics associated to de Sitter (dS) space presents a dilemma in the size of $\Lambda$ and the concern related to the dynamical instability due to quantum corrections \cite{Tsamis1996, Mukhanov1997}.  In addition, dS solutions \cite{Danielsson2018} appear to be in conflict with the weakly coupled and weakly curved string theory.  On the other hand,  Anti-de Sitter (AdS) space is crucial to the theories of supergravity and superstrings \cite{Castellani1991, Witten1998}. 

In this context various models based on AdS space have been proposed. An idea of the transition between a dS space and an AdS space glued adjacent to a bubble wall having self-consistent vacua is prescribed in \cite{Li1999}.  Physically, it leads to the possibility of transitioning to a negative $\Lambda$ or the universe in AdS space. Following the well-known Randall–Sundrum models another work \cite{Barcelo2000} have demonstrated the D-brane (where the observable universe is localised) as a real and physical border to the five-dimensional AdS spacetime. Thus, AdS/CFT conjecture \cite{Maldacena1998}  based on the Randall–Sundrum models can be effective on a boundary of the (asymptotic) AdS space which accords with the observations. Moreover, in a novel work \cite{Banerjee2018} it is suggested that a 4 dimensional observer can perceive  an effective 4 dimensional dS spacetime on restricting  four dimensional gravity  to a brane  which moderates  the decay of five-dimensional AdS false vacuum to supersymmetric true $AdS_5$ vacuum. Additionally, a recent investigation \cite{Henke2021}  presented an AdS  brane model avoiding the quadratical energy density which can explain cosmological observations as similar manner to cold dark matter (CDM) model. Furthermore, an original contemporary research \cite{Biasi2022} considered the development of dS bubbles via processes resembling to the Bizon-Rostworowski instability in the global background of AdS space in general relativity. This paves the way to explore how AdS fluctuations can generate multiple accelerating universes.

Although the stability analysis of pulsars in $f(Q)$ gravity in AdS space experiences less motivation from an observational point of view than that in dS space,  it is an intriguing topic in theoretical physics for various reasons. To begin with,  the proposed AdS/CFT conjecture  \cite{Maldacena1998} in string theory has generated curiosity  in finding  the possibile solutions to space-times with a negative cosmological constant and maximum symmetry \cite{Hawking1973}. Additionally, an investigation \cite{Gazeau2020} which is based on the transition of phases between epochs of quark–gluon plasma and the hadron indicates that the curvature energy in an effective AdS might be the source of  dark matter energy.  Moreover, AdS spacetime being a maximally symmetric spacetime, can be  utilised to develop effective model describing  the interaction between matter field and gravity.  Consequently, solitons for instance boson stars, scalar fields and gravitational interaction iduced compact objects are capable of being sustained in  AdS \cite{Aste2003, Prikas2004, Nogu2013, Brih2013, Buch2013, Kicha2014, Kumar2015, Brih2022}. Again, in vacuum an action which is a subclass of Horndeski gravity and  depicts  a nonminimal derivative coupling of  the Einstein tensor  with a scalar field can generate black hole solutions  in asymptotically AdS with configuration of  nontrivial scalar field \cite{Rinaldi2012, Minamitsuji2014, Anabalon2014, Babichev2015}. Subsequently,  Neutron stars with the action and AdS space have been investigated in a series of works \cite{Cist2015, Cist2016, Maselli2016, Silva2016, Salcedo2018} assuming a vanishing “bare” $\Lambda$ and a vanishing standard kinetic term as special case of the study. However, in a recent  paper \cite{Pavel2023}, with non zero $\Lambda$ and standard kinetic term configurations of neutron star have been studied thoroughly for  a perfect fluid content along with the polytropic equation of state and the action as subclass of Horndeski gravity. Thus, the study \cite{Pavel2023} have shown that the effective negative  $\Lambda_{AdS}$ contributes to form the particular relationship of mass with radius. Finally, a positive $\Lambda$ inevitably modetates gravitational collapse whereas the negative $\Lambda$ is complementary to  the gravitational forces. Above all, positive anisotropic force due to pressure anisotropy being repulsive plays significant role to stabilize the stellar systems in both dS and AdS spacetime.  Based on the current cosmological observations as well as the present progress on the braneworlds related to supergravity and superstring theories, we are inspired to study  the comparative analysis of the stability  of pulsars in both dS and AdS spacetime  in the framework $f(Q)$ gravity.}

The paper is organized as follows: Sec. \ref{sec2} presents a concise overview of the underlying principles of modified $f(Q)$-gravity theory. Moving on to Sec. \ref{sec3}, an exact anisotropic solution in $f(Q)$-gravity is derived. The determination of boundary conditions is addressed in Sec. \ref{sec4}, where the interior spacetime is matched with the exterior spacetime, specifically { the Schwarzschild de Sitter spacetime and Schwarzschild Anti-de Sitter spacetime},  at the pressure-free boundary. Sec. \ref{sec5} delves into the investigation of the physical viability of the stellar model, focusing on the behavior of various thermodynamic variables such as energy density, radial and tangential stresses, and the anisotropic parameter. In Secs. \ref{sec6} and \ref{sec7}, the maximum mass and radii predictions for different neutron stars are discussed, along with stability analysis using the adiabatic index and Harrison-Zel'dovich-Novikov criteria, respectively. Finally, Sec. \ref{sec8} provides concluding remarks on our findings.

{\color{black}\section{Modified $f(Q)$-gravity theory }\label{sec2}

The processes for the modified $f(Q)$ gravity technique are as follows \citep{Zhao:2021zab,Jimenez_2018_98}: the action for $f(Q)$-gravity with matter sources is given by

\begin{eqnarray}
\label{eq1}
\mathcal{S}=\int\frac{1}{2}\,f(Q)\sqrt{-g}\,d^4x+\int \mathcal{L}_m\,\sqrt{-g}\,d^4x,~~~
\end{eqnarray}
where the symbol $\mathcal{L}_m$ denotes the Lagrangian density for matter distribution while $Q$ represents a nonmetricity scalar. The energy-momentum tensor $T_{\epsilon\nu}$ connected to Lagrangian $\mathcal{L}_m$ is expressed as 
\begin{eqnarray}
\label{eq2}
&& \frac{2}{\sqrt{-g}}\frac{\delta\left(\sqrt{-g}\,\mathcal{L}_m\right)}{\delta g^{\epsilon\nu}}={T}_{\epsilon\nu}.
\end{eqnarray}
The tensor $Q_{\lambda\epsilon\nu}$ for nonmetricity term is calculated as 
\begin{equation}\label{eq4}
Q_{\lambda\epsilon\nu}=\bigtriangledown_{\lambda} g_{\epsilon\nu}=\partial_\lambda g_{\epsilon\nu}-\Gamma^\delta_{\,\,\,\lambda \epsilon}g_{\delta \nu}-\Gamma^\delta_{\,\,\,\lambda \nu}g_{\epsilon \delta},
\end{equation}
where, $\Gamma^\delta_{\,\,\,\epsilon\nu}$ defines the affine connection, further it is defined as
\begin{equation}\label{eq5}
\Gamma^\delta_{\,\,\,\epsilon\nu}=K^\delta_{\,\,\,\epsilon\nu}+ L^\delta_{\,\,\,\epsilon\nu}+\lbrace^\delta_{\,\,\,\epsilon\nu} \rbrace,
\end{equation}
with
\begin{eqnarray}
\label{eq6}
&&\hspace{0.3cm}  \frac{1}{2} T^\delta_{\,\,\,\epsilon\nu}+T_{(\epsilon\,\,\,\,\,\,\nu)}^{\,\,\,\,\,\,\delta}=K^\delta_{\,\,\,\epsilon\nu},\nonumber\\
&&\hspace{0.2cm}
\frac{1}{2}Q^\delta_{\,\,\,\epsilon\nu}-Q_{(\epsilon\,\,\,\,\,\,\nu)}^{\,\,\,\,\,\,\delta}=L^\delta_{\,\,\,\epsilon\nu},\nonumber \\
&&\hspace{0.2cm} \frac{1}{2}g^{\delta\sigma}\left(\partial_\epsilon g_{\sigma\nu}+\partial_\nu g_{\sigma\epsilon}-\partial_\sigma g_{\epsilon\nu}\right)=\lbrace^\delta_{\,\,\,\epsilon\nu} \rbrace,
\end{eqnarray}
where $L^\delta_{\,\,\,\epsilon\nu}$,  $\lbrace^\delta_{\,\,\,\epsilon\nu} \rbrace$, $T^\delta_{\,\,\,\epsilon\nu}$, and $K^\delta_{\,\,\,\epsilon\nu}$ are known as the disformation, Levi-Civita connection, torsion tensor and the contortion tensors respectively. Further, the anti-symmetric part of the affine connection can be defined as $T^\delta_{\,\,\,\epsilon\nu}=2\Gamma^\lambda_{\,\,\,[\epsilon\nu]}$. This last expression also described by the torsion tensor $T^\delta_{\,\,\,\epsilon\nu}$. 

Finally, the non-metricity scalar expression can be cast as
\begin{equation}\label{eq9}
-Q_{\alpha\epsilon\nu}\,P^{\alpha\epsilon\nu}=Q. 
\end{equation}
The above equation $P^{\alpha\epsilon\nu}$ provides a non-metricity conjugate. The corresponding tensor is defined as
\begin{equation}\label{eq7}
\frac{1}{4}\left[-Q^\alpha_{\,\,\epsilon\nu}+2 Q_{(\epsilon\,\nu)}^\alpha+Q^\alpha g_{\epsilon\nu}-\tilde{Q}^\alpha g_{\epsilon\nu}-\delta^\alpha_{(\epsilon}Q_{\nu)}\right]=P^\alpha_{\,\,\epsilon\nu}.
\end{equation}
Here, $\tilde{Q}_\alpha$ and $Q_{\alpha}$ are the two independent traces.  Both the traces can be defined by the following relation
\begin{equation}\label{eq8}
\tilde{Q}_\alpha=Q^\epsilon_{\,\,\,\,\alpha\epsilon}, \;\;\;Q_{\alpha}\equiv Q_{\alpha\,\,\,\epsilon}^{\,\,\epsilon}.
\end{equation}

To get the appropriate field equations for $f(Q)$-gravity, it is required to vary the action (\ref{eq1}) with respect to the metric tensor $g^{\epsilon\nu}$. Thus, one can obtain the $f(Q)$ gravity field equations as follows: 
\begin{eqnarray}
\label{eq10}
&& \hspace{.01cm}T_{\epsilon\nu}=\frac{2}{\sqrt{-g}}\bigtriangledown_\gamma\left(\sqrt{-g}\,f_Q\,P^\gamma_{\,\,\,\,\epsilon\nu}\right)+\frac{1}{2}g_{\epsilon\nu}f 
+f_Q\big(P_{\epsilon\gamma i}\,Q_\nu^{\,\,\,\gamma i}-2\,Q_{\gamma i \epsilon}\,P^{\gamma i}_{\,\,\,\nu}\big),~~
\end{eqnarray}
where $f_Q=\frac{d f}{d Q}$. In addition, Eq. (\ref{eq1}) can be used to construct an extra connection constraint. Consequently, It is defined as
\begin{equation}\label{eq11}
\bigtriangledown_\epsilon \bigtriangledown_\nu \left(\sqrt{-g}\,f_Q\,P_\gamma^{\,\,\,\,\epsilon\nu}\right)=0.
\end{equation}

Other restrictions, particularly torsion-free and curvature-free, permits us to represent the affine connection as 
\begin{equation}\label{eq12}
\Gamma^\lambda_{\,\,\,\epsilon\nu}=\left(\frac{\partial x^\lambda}{\partial\xi^\beta}\right)\partial_\epsilon \partial_\nu \xi^\beta.
\end{equation}
where, $\Gamma^\lambda_{\,\,\,\epsilon\nu}=0$ can be obtained for coincident gauge. As a result, from Eq. (\ref{eq4}), we have the following expression
\begin{equation}\label{eq13}
Q_{\lambda\epsilon\nu}=\partial_\lambda \,g_{\epsilon\nu}.
\end{equation}

On the other hand, we consider the spherically symmetric metric of the form, 
\begin{eqnarray}\label{eq14}
ds^2=-e^{Y(r)}dt^2+e^{X(r)}dr^2+r^2 d\theta^2+r^2 \sin^2\theta \,d\phi^2,
\end{eqnarray}
where $Y(r)$ and $X(r)$ are functions of $r$ only. \\

The non-metricity scalar $Q$ for the spacetime (\ref{eq14}) is given by,  
\begin{eqnarray}\label{eq15}
Q=\frac{-2  \left(r Y^\prime(r)+1\right)}{r^2 e^{X(r)}}.
\end{eqnarray}

Furthermore, we consider the matter distribution to be anisotropic. In this case, the source $T_{\nu \epsilon}$ is given as, 
\begin{eqnarray}
 T_{\epsilon \nu} = \rho\, \eta_{\epsilon } \eta_{\nu}  + P_t (\eta_{\epsilon } \eta_{\nu} +g_{ \epsilon \nu }-\chi_{\epsilon}\,\chi_{\nu})+P_r \chi_{\epsilon}\,\chi_{\nu}.  \label{em}
\end{eqnarray}
Then the non-zero components of the energy-momentum tensor are
\begin{eqnarray}
&& T^0_0= -\rho, ~~~~ T^1_1= P_r,~~~T^2_2=T^3_3 = P_t.~~~~~~~\label{eq16}
\end{eqnarray}

For the spherically symmetric metric (3), the final form of the field equations (assuming that the affine connection is zero) can be written as,   
\begin{eqnarray}
&& \hspace{-1.0cm}\rho =-f_{Q}\Big[Q+\frac{1}{r^2}+\frac{1}{r\,e^{X}}(Y^\prime+X^\prime)\Big]+\frac{f(Q)}{2},\label{eq17}\\
&& \hspace{-1.0cm} P_r=f_{Q}\Big[Q+\frac{1}{r^2}\Big]-\frac{f(Q)}{2},\label{eq18}\\
&& \hspace{-1.0cm} P_t=f_{Q}\Big[\frac{Q}{2}-\frac{1}{e^{X}} \Big\{\frac{Y^{\prime \prime}}{2}+\frac{(r Y^\prime+2)}{4r}  (Y^\prime-X^\prime)\Big\}\Big]-\frac{f(Q)}{2},~~~~~\label{eq19}\\
&& \hspace{-1.0cm} 0=\frac{\text{cot}\,\theta}{2}\,Q^\prime\,f_{QQ}. \label{eq20}
\end{eqnarray}

In order to proceed with obtaining models of compact stellar objects, we assume that the interior matter distribution, under the influence of $f(Q)$ gravity, is endowed with anisotropic pressure.  The results obtained by  Zhao~\cite{Zhao:2021zab} on the compatibility of static SS spacetimes with the coincident gauge are very significant. In light of Zhao's findings, with the affine connection being zero in the specified coordinate framework, and further enforcing the requirement that the $f(Q)$-gravity theory yields vacuum solutions, particular $T_{\epsilon \nu}=0$, it is possible to deduce that the off-diagonal component as given by Eq. (\ref{eq20}), 
 \begin{eqnarray}
   \frac{\text{cot}\,\theta}{2}\,Q^\prime\,f_{QQ}=0, \label{eq16a}
\end{eqnarray}
where, $Q$ is given by Eq.~(\ref{eq15}). 

Furthermore, the equations of motion, when combined with the diagonal components (\ref{eq16a}), lead to $f_{QQ}$ being zero. This finding suggests that choosing $f(Q)=Q^2$ initially will lead to an inconsistent system of equations of motion. If the function $f(Q)$ is not linear with respect to $Q$, then the metric equation (\ref{eq14}) combined with an affine connection $\Gamma^{\lambda}_{\epsilon \nu}=0$ does not provide an acceptable solution for the equations of motion. The nonexistence of static SS vacuum solutions in $f(Q)$ theory does not always indicate the absence of such solutions. Instead, it implies that the SS coordinate system does not align with the coincident gauge. To get SS solutions, one needs to think about a more extensive definition of the static SS metric. For a comprehensive understanding of this aspect and to get a detailed explanation, please refer to Ref.~\citep{Zhao:2021zab}. 

 Moreover, the research carried out by Avik \& Loo~\cite{Avik} indicates that the presence of classical events cannot be guaranteed within the framework of the symmetric teleparallel theory. However, it relies on the distinct attributes of the $f(Q)$ model. The authors give further evidence that supports the correspondence between the energy conservation criteria and the field equation of the affine connection within the framework of $f(Q)$ theory. It is crucial to note that, except the linear form of $f(Q)$, the non-linear $f(Q)$ models fail to satisfy the energy conservation criterion or, equivalently, the ensuing field equation for all spacetime geometries, as long as the variable $Q$ stays constant. 
In this regard, Wang et al.~\cite{Wang} also proved that the exact Schwarzschild solution is only valid when the functional form of $f(Q)$ is linear, and they have imposed constraints on its derivation and limitation. Together with a deep-dive into related work in the existing literature and the compatibility of the proposed solution, we have assumed 
\begin{eqnarray}
f_{QQ}=0~\Longrightarrow~  f_{Q}=-\alpha~  ~\Longrightarrow~ f(Q)=-\alpha\,Q-\beta ,~~~  \label{eq21}
\end{eqnarray}
where $\alpha$ and $\beta$ are constants. {Note that in the case of constant $Q$ this framework is identical to GR, with a cosmological constant.}\\
The set of equations for motion that results when Eqs. (\ref{eq14}) and (\ref{eq21}) are replaced into Eqs. (\ref{eq17}) - (\ref{eq19}) are as follows:
\begin{eqnarray}
&& \hspace{-1.0cm}\rho = \frac{1}{2\,r^2} \Big[2\, \alpha+2\, e^{-X}\, \alpha  \left(r\, X^\prime-1\right)-r^2 \,\beta \Big],\label{eq22}\\
&& \hspace{-1.0cm} P_r=\frac{1}{2\,r^2} \Big[-2\, \alpha+2\, e^{-X} \,\alpha  \left(r\, Y^\prime+1\right)+r^2\, \beta\Big],\label{eq23}\\
&& \hspace{-1.0cm} P_t=\frac{e^{-X}}{4\,r} \Big[2\, e^{X}\, r\, \beta -\alpha\,  \left(2+r Y^\prime\,\right) \left(X^\prime-Y^\prime\right) +2\, r\, \alpha \, Y^{\prime \prime}\Big].~~~~~~\label{eq24} 
\end{eqnarray}
The mass function under $f(Q)$-gravity regime can be calculated with the aid of the  formula,
\begin{eqnarray}
m(r)=\frac{1}{2} \int_0^r x^2\,\rho(x)  \,dx .
\end{eqnarray}
The linear combination of (\ref{eq22})-(\ref{eq24}) yields, 
\begin{eqnarray}
&&   -\frac{Y^\prime}{2}(\rho+P_r)-P^{\prime}_r+\frac{2}{r} (P_t-P_r)=0.~~~\label{eq25}
\end{eqnarray}
Surprisingly the equation (\ref{eq25}) corresponds to the conservation formula in GR and is often referred to as the TOV equation in the $f(Q)$-gravity theory \citep{TOV1,TOV2,Wang}. 
Our objective is to find an exact solution to the $f(Q)$-gravity field equations (\ref{eq22})-(\ref{eq24}) by considering an anisotropic distribution of matter which we provide in the next section. 

\section{Exact anisotropic solution in $f(Q)$-gravity}
\label{sec3}
As we can see that we have five unknowns: $\rho(r)$, $P_r(r)$, $P_t(r)$, $Y(r)$, and $X(r)$ and three independent equations. It is very important to mention that various techniques, including the MIT bag model or polytropic EOS, Karmarkar conditions, and complexity factor conditions, have been used to solve the system of equations in $f(Q)$-gravity along with various modified gravity theories. On the other hand, Harko and his collaborators~\citep{Harko2002,Mak2002,Mak2003} investigated the most general solution of Einstein's gravitational field equations describing spherically symmetric and static anisotropic stellar type configurations using particular form of anisotropy and density profile. Therefore, we use the same methodology to solve the $f(Q)$-gravity system to determine the most general exact solution for an anisotropic distribution of matter. To achieve this objective, we use the anisotropic condition in equations (\ref{eq23}) and (\ref{eq24}), which is derived by subtracting equation (\ref{eq18}) from equation (\ref{eq19}). The pressure anisotropy condition in $f(Q)$-gravity is obtained in the following manner: 
\begin{eqnarray}
\alpha\,\big[ (2 Y^{\prime \prime}+ Y^{\prime 2}) r^2-2  Y^{\prime} r- X^{\prime} r ( Y^{\prime}r+2)+4 (e^{X}-1)\big] -4 r^2 \Delta\,e^{X}=0.~~\label{eq26}
\end{eqnarray}
The above equation contains three unknowns namely $Y(r)$, $X(r)$, and $\Delta(r)$.  

To solve the master equation (\ref{eq26}),  we shall consider the following physically valid expression for variable $X(r)$ given by the Durgapal-Fuloria ansatz~\citep{DF}: 

\begin{equation}
X(r)=\ln\Bigg[\frac{7+14Ar^2+7A^2r^4}{7-10Ar^2-A^2r^4}\Bigg], \label{eq28}
\end{equation} where $A$ is a constant. The Durgapal-Fuloria ansatz has been widely used to model superdense stars in classical general relativity as the resulting stellar models display various salient features with regards to regularity, stability and causality. It was shown that the ensuing models could have masses $\geq 3.73$ $M_{\odot}$ which makes it suitable to model the secondary component of gravitational events such as GW190814. Using Eq. (\ref{eq28}), the anisotropy condition (15) along with transformation $Y=2\ln\Psi$ reduce to 
\begin{eqnarray}
&& \hspace{-0.5cm}\frac{d^2\Psi}{dr^2}-\Bigg[\frac{A^3 r^6+19 A^2 r^4-21 A r^2-7}{r \left(A^3 r^6+11 A^2 r^4+3 A r^2-7\right)}\Bigg] \frac{d\Psi}{dr}-\frac{8 \alpha A^2 r^3 \left(A r^2+5\right) -7 \Delta  r \left(A r^2+1\right)^3}{r \,\alpha\,\left(A^3 r^6+11 A^2 r^4+3 A r^2-7\right)} \Psi=0.~~~~~~~ \label{eq29}
\end{eqnarray}

Since the above equation involves two undetermined unknowns $\Psi$ and $\Delta$, we have to specify a viable expression for $\Delta$ which yields the closed-form solution which we seek. Employing physical considerations based on regularity together with some mathematical manipulation, we conceive the following expression for $\Delta$,

\begin{equation}\label{eq30}
\Delta={\frac {8\alpha \Delta_0\,A^2r^{2}\big[2(1-8Ar^2-A^2r^4)+\Delta_0(Ar^2 +5)\big]}{7  \left( 1+Ar^2
 \right) ^{3} \left( 1-\Delta_0+Ar^2 \right) ^{2}}}. 
\end{equation}

Here $\Delta_0$ is a positive constant.  To ensure positive anisotropy throughout the model, the factor: $\big[2(1-8Ar^2-A^2r^4)-\Delta_0(Ar^2 -5)\big]$, must be positive. This condition gives the constraint as:
\begin{eqnarray}
    0 < Ar^2 <
\frac{\sqrt{{\Delta_0}^{2}+8\Delta_0+272}-(16-\Delta_0)}{4}.
\end{eqnarray}
By plugging the expression of $\Delta$ in Eq.(\ref{eq29}), we get a general second order ODE in $\Psi$,
\begin{eqnarray}
&& \hspace{-0.6cm}\frac{d^2\Psi}{dr^2}-\Bigg[\frac{A^3 r^6+19 A^2 r^4-21 A r^2-7}{r \left(A^3 r^6+11 A^2 r^4+3 A r^2-7\right)}\Bigg] \frac{d\Psi}{dr}-\frac{8 A^2 r^2 \left(\Psi_1(r)-\Delta _0^2 \left(A r^2+5\right)+6 A r^2+5\right)}{\left(A^3 r^6+11 A^2 r^4+3 A r^2-7\right) \left(A r^2-\Delta _0+1\right)} \Psi=0,~~~~~~~ \label{eq32}
\end{eqnarray}
where, $\Psi_1(r)=A^2 r^4+\Delta _0 \left(2 A^2 r^4+15 A r^2-7\right)$. If we look closely at Eq. (\ref{eq32}), we note that setting: 
$\Psi=(1-\Delta_0+Ar^2)^2$ satisfies the above equation. \\

Then the most general solution can be found by assuming:
$\Psi=(1-\Delta_0+Ar^2)^2$ as a particular solution of Eq. (\ref{eq32}). Let us assume, $\Psi=(1-\Delta_0+Ar^2)^2\,\Omega$ is a general solution of Eq. (\ref{eq32}). We then obtain 
\begin{eqnarray}
&&\hspace{-0.6cm}\Omega(r)=  \Bigg[\tilde{C}\int\exp\Bigg\{\int\Bigg(\frac{7-C^{3}r^{6}
-19A^2r^{4}+21Ar^{2}}{r\left(A^3 r^6+11 A^2 r^4+3 A r^2-7\right)}
-\frac{4Ar^{2}}{r(1-\Delta_0+Ar^2)}\Bigg)dr\Bigg\}dr+\tilde{D}\Bigg], ~~~~~~
\end{eqnarray}
where, $\tilde{C}$ and $\tilde{D}$ are arbitrary constants of integration. After integration of the above equation, we get 
\begin{eqnarray}
&& \hspace{-0.1cm}\Psi=\Bigg[C\Bigg\{\frac{[{\Delta_{01}}+{\Delta_{0 2}}(1-\Delta_0+Ar^2)+{\Delta_{0 3}}\Omega_3(r)]\Omega_1(r)}{(1-\Delta_0+Ar^2)^3} +\Omega_2(r)\Bigg\}+D\Bigg]\Omega_3(r),~\label{eq3.8}
\end{eqnarray}
where, new arbitrary constants $C$ and $D$ have been introduced while 
\begin{eqnarray}
&& \hspace{-0.5cm}  \Omega_1(r)=\sqrt{({\Delta_{0 5}}-2(4+\Delta_0)(1-\Delta_0+Ar^2)
-\Omega_3(r))}, ~~~~ \Omega_2(r)=\frac{{\Delta_{0 4}}}{\sqrt{{\Delta_{0 5}}}}
\ln\left[\frac{\Omega_4(r)+
\sqrt{{\Delta_{0 5}}}\Omega_1(r)}{(1-\Delta_0+Ar^2){\Delta_{0 5}}}\right],\nonumber\\
&& \hspace{-0.5cm} \Omega_3(r)=(1-\Delta_0+Ar^2)^2,~~~~\Omega_4(r)=\Delta_{0 5}-(4+\Delta_0)(1-\Delta_0+Ar^{2}),~~
{\Delta_{0 1}}=\frac{\Delta_0}{3(16-8\Delta_0-\Delta_0^{2})},\nonumber\\
&& \hspace{-0.5cm}
{\Delta_{02}}=\frac{24-2\Delta_0+\Delta_0^{2}}{3(16-8\Delta_0-\Delta_0^{2})^{2}},~~~~~
{\Delta_{03}}=\frac{288+80\Delta_0-10\Delta_0^{2}+\Delta_0^{3}}{3(16-8\Delta_0-\Delta_0^{2})^{3}},\nonumber\\
&& \hspace{-0.5cm}{\Delta_{04}}=\frac{1536-384\Delta_0+48\Delta_0^{2}-2\Delta_0^{3}}{3(16-8\Delta_0-\Delta_0^{2})^{3}},~~
{\Delta_{0 5}}=(16-8\Delta_0-\Delta_0^{2}). \nonumber
\end{eqnarray}
The expressions for energy density and pressures may be derived using equations (\ref{eq22})-(\ref{eq24}) along with Eqs. (\ref{eq28}) and (\ref{eq3.8}) as,  
\begin{eqnarray}
&&\rho=\frac{8A\alpha(9+2Ar^{2}+A^2r^{4})}{7(1+Ar^{2})^{3}} -\frac{\beta}{2}, \\
&& P_r= \frac{1}{7 r \Psi(r) \left(A r^2+1\right)^2}\Big\{2 \alpha  \big[\Psi_{11}(r) \left(7-A r^2 \left(A r^2+10\right)\right)-4 A r \Psi(r) \left(A r^2+3\right)\big]\Big\}+\frac{\beta }{2},~~~~~~~~\\
&& P_t=\frac{1}{14 r (1 + A r^2)^3 \Psi(r)} \Big[ -2 \alpha  \Psi_{22} (r) r \big(A^3 r^6+11 A^2 r^4 +3 A r^2-7\big)-2 \alpha  \Psi_{11}(r) \nonumber\\
&&\hspace{0.7cm} \times \left(A^3 r^6+3 A^2 r^4+27 A r^2-7\right)+r \Psi(r) \big\{7 A^3 \beta  r^6+A^2 \left(21 \beta  r^4+16 \alpha  r^2\right)+A \big(21 \beta  r^2 \nonumber\\
&&\hspace{0.7cm} -48 \alpha \big)+7 \beta \big\}\Big]
\end{eqnarray}
where, $\Psi_{11}(r)=\frac{d\Psi}{dr}$,~~~$\Psi_{22}(r)=\frac{d^2\Psi}{dr^2}$.

\section{Boundary Condition}\label{sec4}
The junction conditions are an essential part of bounded astrophysical stellar systems, which could be determined by matching the interior spacetime with the exterior spacetime at the pressure-free boundary. { In the linear $f(Q)$-gravity, we have considered a suitable exterior spacetime given as,}
\begin{eqnarray} \label{eq36}
&& \hspace{-0.3cm} ds^2_+ =-\bigg(1-\frac{2M}{r}-\frac{\Lambda}{3}~r^2\bigg)\,dt^2+\frac{dr^2}{1-\frac{2M}{r}-\frac{\Lambda}{3}~r^2} +r^2 \Big(d\theta^2  +\sin^2\theta\,d\phi^2 \Big),
\end{eqnarray} 
where $\Lambda$ indicates a cosmological constant ($\Lambda=\beta/2\alpha$) while $M$ is the total mass of object with $R$ and $M={m(R)}/{\alpha}$. { For positive $\Lambda$ the spacetime will be Schwarzschild de Sitter (dS) and for negative $\Lambda$ it will be Schwarzschild Anti-de Sitter (AdS).} For smooth matching, we use the Israel-Darmois junction conditions, which are mainly: (i) Smooth matching of metric functions at the boundary $r=R$), (ii) vanishing of radial pressure at the stellar boundary $r=R$. Mathematically, we can express these as: 
\begin{eqnarray}
 \bigg(1-\frac{2M}{R}-\frac{\Lambda}{3}~R^2\bigg) &=& e^{Y(R)},~~~\label{eq37}\\
 \bigg(1-\frac{2M}{R}-\frac{\Lambda}{3}~R^2\bigg) &=& e^{-X(R)},~~~\label{eq38}\\
 P_r(R) &=& 0.  \label{eq39}
\end{eqnarray}
The constants $C$ and $D$ will be obtained via the above-mentioned matching conditions.

\section{Physical Analysis}\label{sec5}
In this section, we investigate the physical viability of our stellar model in terms of the behaviours of the thermodynamical variables, viz., energy density, radial and tangential stresses and the anisotropic parameter. An inspection of Figure \ref{f1}, reveals that the energy density is regular at each interior point of the configuration. The energy density is a monotonically decreasing function of the radial coordinate, attaining a maximum value at the centre of the self-gravitating object. The key difference in the left and right panels of Figure \ref{f1} is the change in magnitude of $\alpha$ while the anisotropy factor, $\Delta_0$ is fixed. We observe that an increase in the magnitude of $\alpha$ leads to a higher core density. This increase accentuates the density in the central regions of the star, ie., there is a 'squeezing' of more matter in central, concentric shells. As one moves away from the centre, towards the surface layers of the star, a change in $\alpha$ has no perceptible effect on the stellar density. 

We now turn our attention to Figure \ref{fig4} in which we have plotted the radial pressure as a function of the radial coordinate. The left two panels display the variation of the radial pressure for $\alpha = 0.5$ and $\alpha = 1$ respectively with the anisotropy parameter, $\Delta_0$ held constant. As with the trend in the density, we note that the radial pressure is continuous everywhere inside the star and decreases monotonically with increasing radial coordinate. In the right two panels, we have varied the anisotropy parameter (left image we set $\Delta_0 = 0.1$ and on the extreme right $\Delta_0 = 0.35$) while $\alpha$ is held fixed. For $\Delta_0 = 0.1$ we observe that the radial pressure in the central region is higher than in the case of $\Delta_0 = 0.35$ by an order of $10^2$. An increase in the anisotropic factor seems to relax the fluid particles with this property being more pronounced in the central regions of the star. 

The trend in the anisotropy parameter, $\Delta(r)$ is portrayed in Figure \ref{fig5}. We recall that $\Delta(r) = P_r - P_t$, with $\Delta(r) > 0$ signifying a repulsive force due to radial stresses dominating the transverse stresses within the stellar configuration. This repulsive force helps stabilise the compact object against the inwardly directed gravitational force. In the left two panels of Figure \ref{fig5}, we have varied $\alpha$ while holding $\Delta_0$ constant. The anisotropy is well-behaved and vanishes at the centre of the configuration. Anisotropy grows in magnitude as one moves outwards towards the boundary which culminates in greater stability of the surface layers of the star. We note that an increase of $\alpha$ from $0.5$ to $1.0$, results in the doubling of the anisotropy within the stellar body. The two right figures in Figure \ref{fig5} show the variation of $\Delta(r)$ with a variation in $\Delta_0$ while $\alpha$ is fixed. As $\Delta_0$ increases, we observe a factor of $10$ increase in anisotropy at each interior point of the gravitating body. The role of $\Delta_0$ is to render the stellar fluid more anisotropic, ie. there is a larger deviation between the radial and tangential stresses which helps stabilise the star. Effectively, $\Delta_0$ can be a measure of physical processes such as phase transitions, neutrino and electron transport within the stellar interior as well as dissipation.

\begin{figure}
\centering
\includegraphics[height=5.9cm,width=15cm]{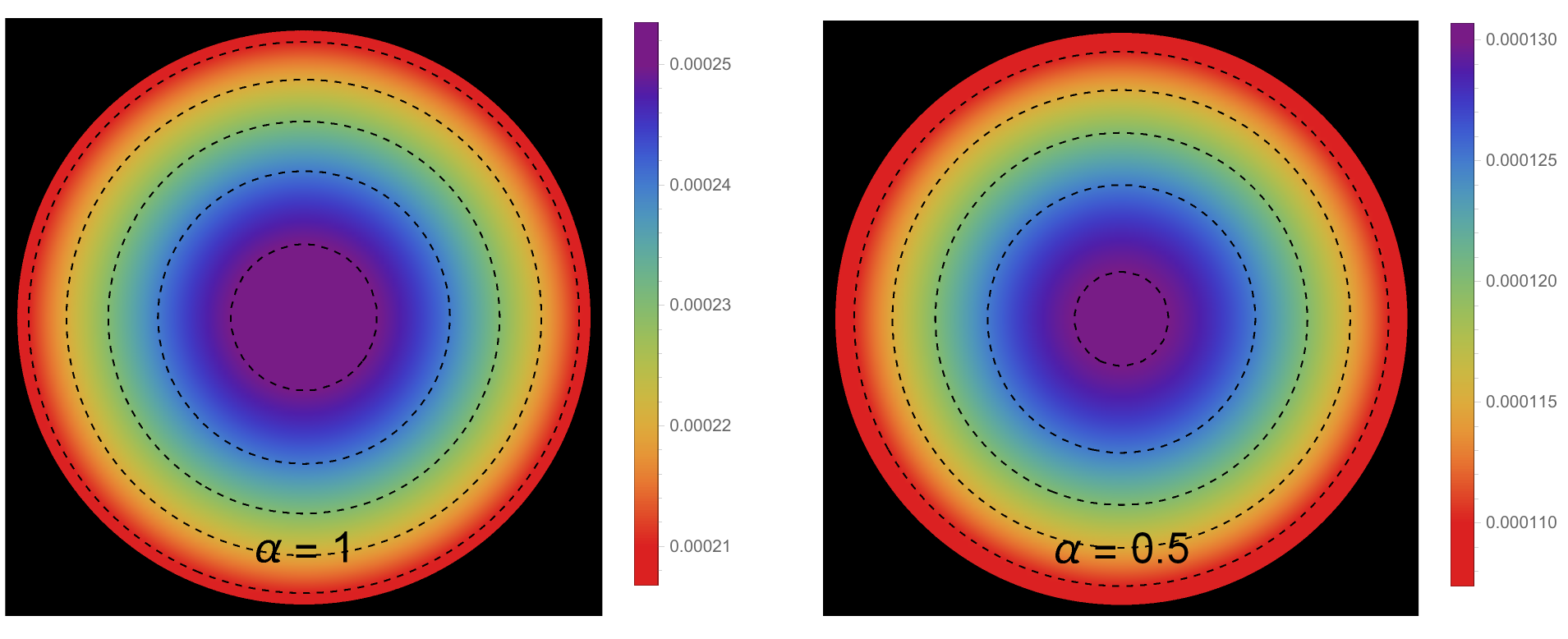}
\caption{The distribution of energy density ($\rho(r)$) in $\text{km}^{-2}$ versus radial distance $r$ from center to boundary of star for taking $A= 0.0006~\text{km}^{-2}$, R = 12.5~$\text{km}$, $\Delta_0 = 0.35$, $\beta = -0.0004~\text{km}^{-2}$.}
\label{f1}
\end{figure}
\begin{figure}
     \centering
      \centering
  \includegraphics[height=5.6cm,width=5.7cm]{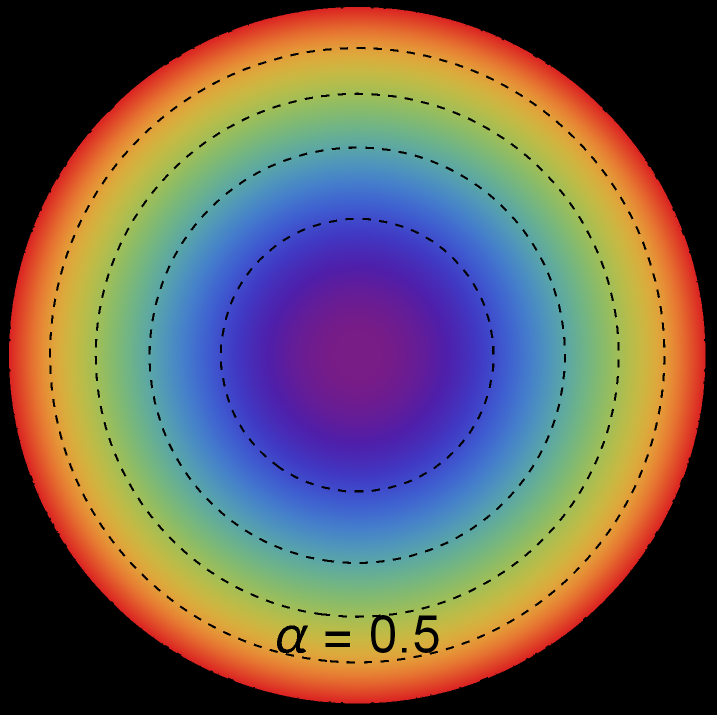}\,\includegraphics[width=1.5cm,height=5.5cm]{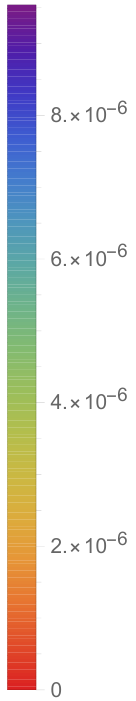}~~~\includegraphics[height=5.6cm,width=5.7cm]{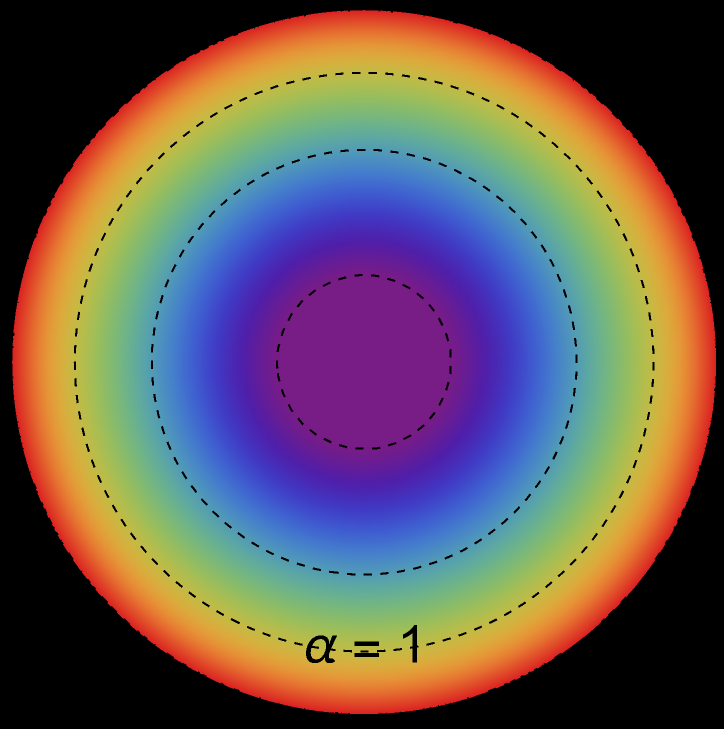}~~~\includegraphics[width=1.5cm,height=5.5cm]{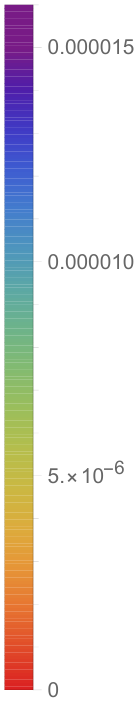}\\
  \vspace{0.5cm}
      \includegraphics[height=5.6cm,width=5.7cm]{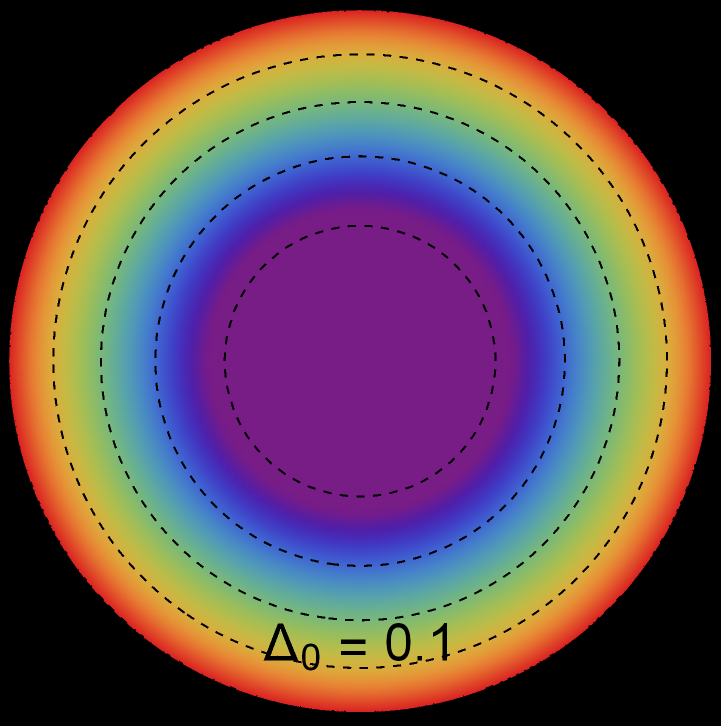}\,\includegraphics[width=1.5cm,height=5.5cm]{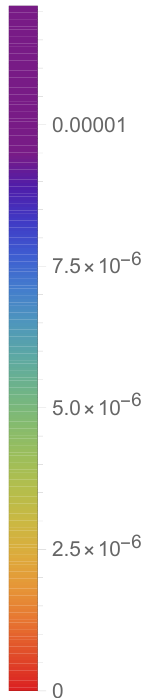} ~~~\,\includegraphics[height=5.6cm,width=5.7cm]{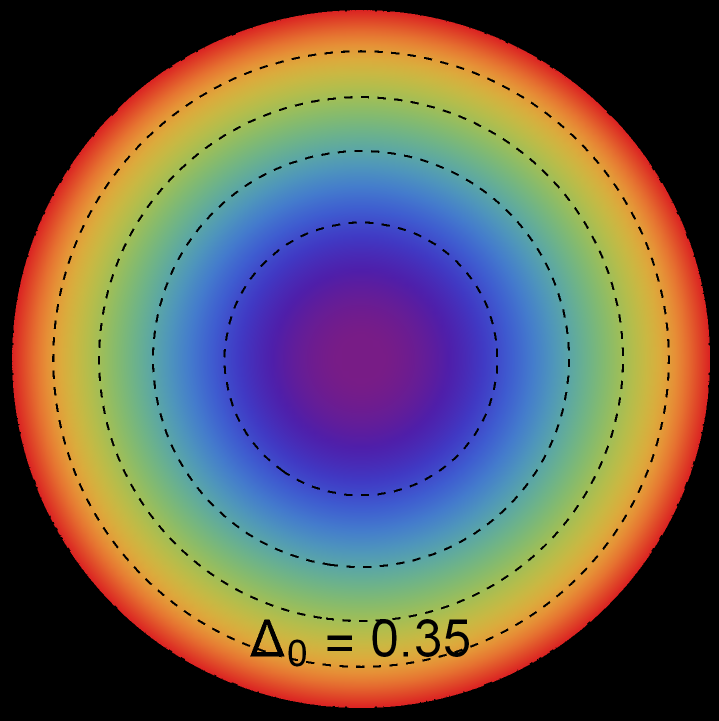}\,\includegraphics[width=1.5cm,height=5.5cm]{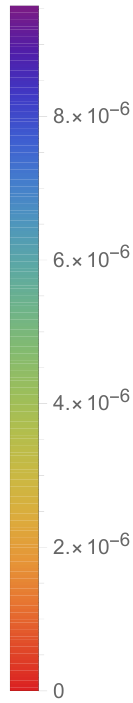}
    \caption{The distribution of radial pressure [$P_r(r)$] in $\text{km}^{-2}$ versus radial distance $r$ from center to boundary of star for different values of $f(Q)$-gravity parameter $\alpha=1$ and $\alpha=0.5$  with fixed anisotropy parameter $\Delta_0=0.35$ (Top panels) while bottom panels represent for anisotropy parameter $\Delta_0=0.35$ and $\Delta_0=0.1$ with fixed $\alpha=0.5$ with the same fixed values as used in Fig.~\ref{f1}.}
    \label{fig4}
\end{figure}
\begin{figure}
     \centering
  \centering
 \includegraphics[height=5.6cm,width=5.7cm]{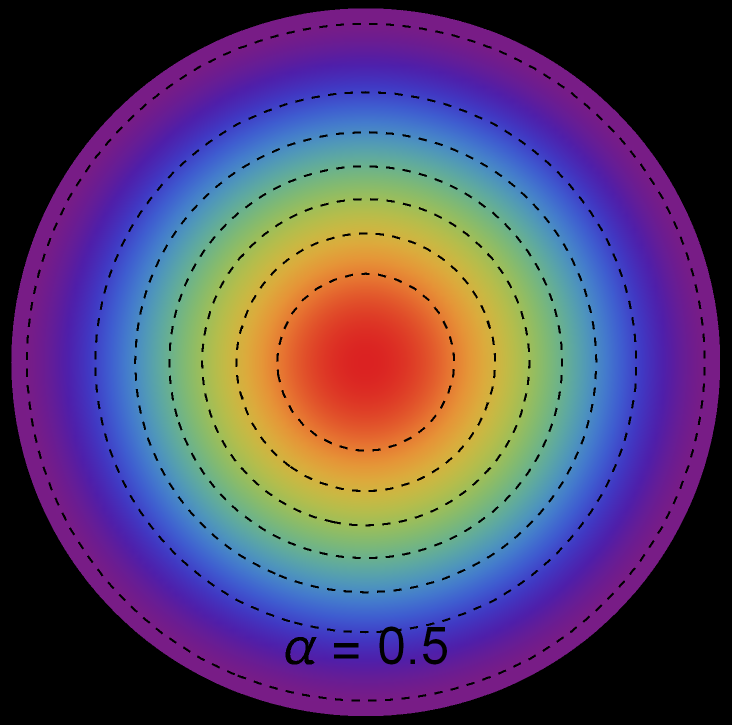}\,\includegraphics[height=5.5cm,width=1.5cm]{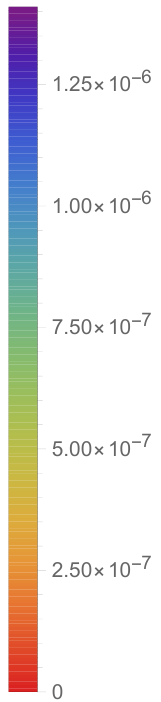}~~~\includegraphics[height=5.6cm,width=5.7cm]{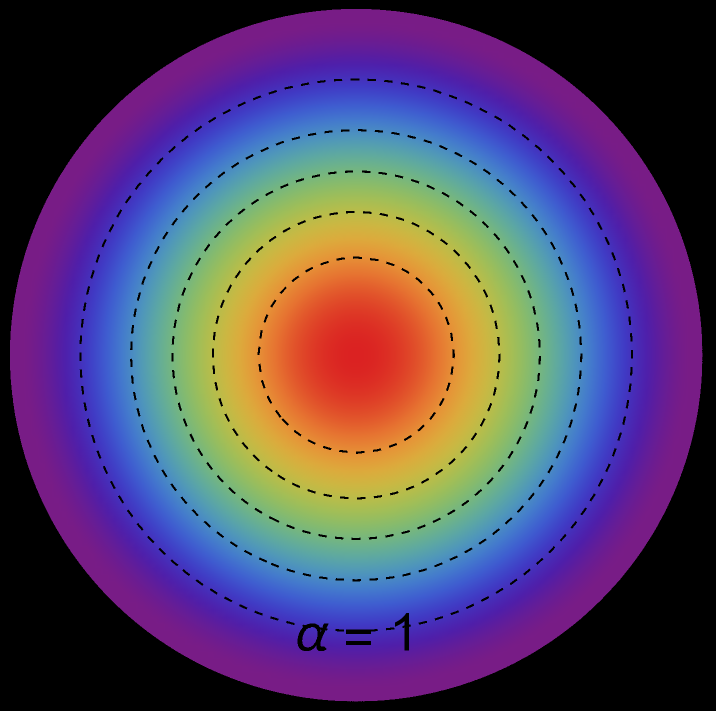}~~~\includegraphics[height=5.5cm,width=1.5cm]{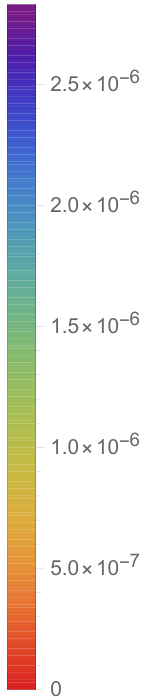}\\
 \vspace{0.5cm}
      \includegraphics[height=5.6cm,width=5.7cm]{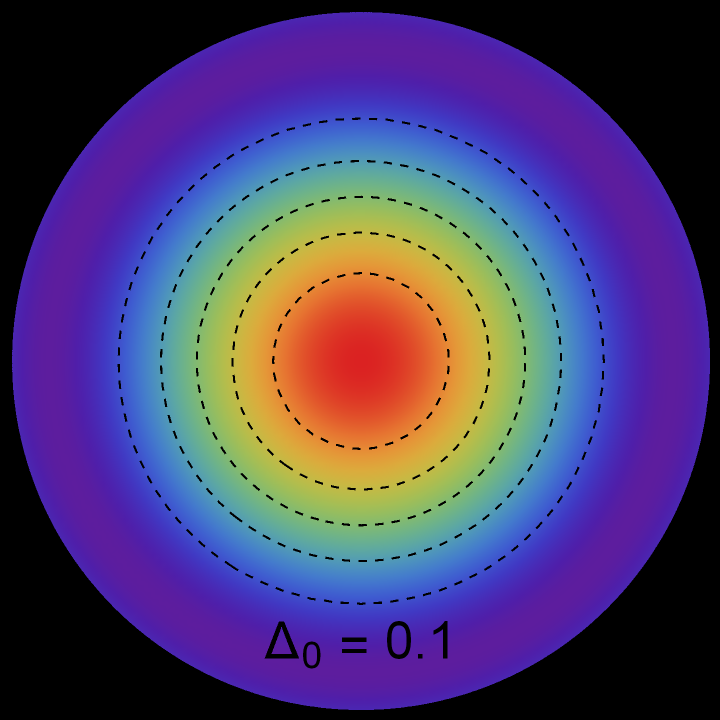}\,\includegraphics[height=5.5cm,width=1.5cm]{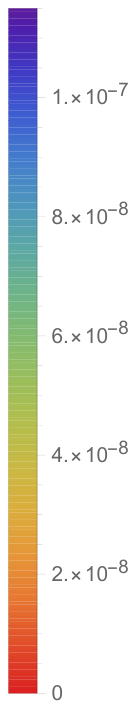} ~~~\,\includegraphics[height=5.6cm,width=5.7cm]{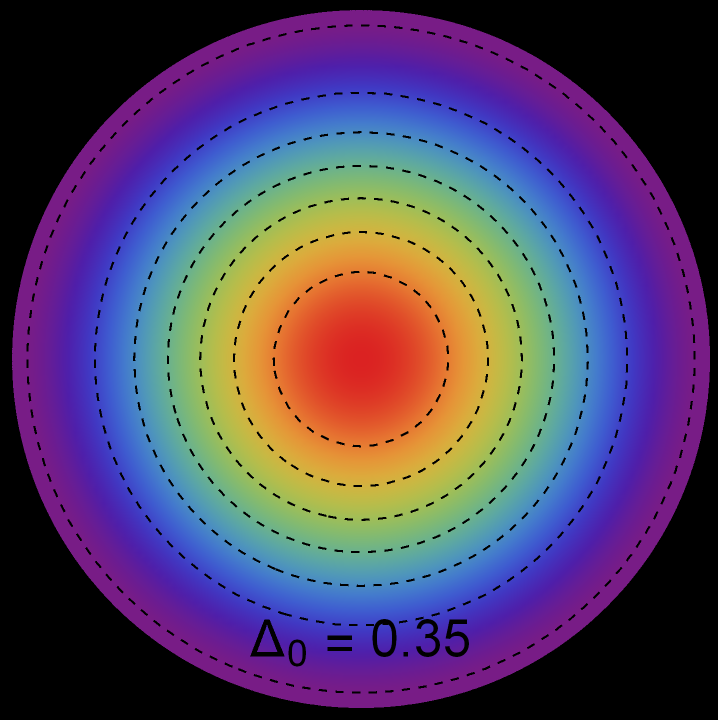}\,\includegraphics[height=5.5cm,width=1.5cm]{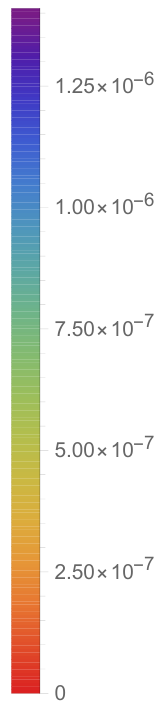}
    \caption{The distribution of anisotropy ($\Delta(r)$) in $\text{km}^{-2}$ versus radial distance $r$ from center to boundary of star for different values of $f(Q)$-gravity parameter $\alpha=1$ and $\alpha=0.5$  with fixed anisotropy parameter $\Delta_0=0.35$ (top two panels) while bottom two panels represent for anisotropy parameter $\Delta_0=0.35$ and $\Delta_0=0.1$ with fixed $\alpha=0.5$. We used the following set of numerical values to plot this figure: $A= 0.0006~\text{km}^{-2}$, R = 12.5~$\text{km}$, $\beta = -0.0004~\text{km}^{-2}$.}
    \label{fig5}
\end{figure}

\section{Prediction of maximum mass and radii for different Neutron stars}\label{sec6}
Pulsars are neutron stars that can emit periodic and strong electromagnetic signals as pulses arising from their high magnetic fields and rotational property.  Large rotational frequencies of such pulsars indicate that they should be highly compact. In general, pulsars can be found in binary systems where the massive neutron star can have one companion star.  These binary systems subject to Kepler’s third law can provide measurements of mass \citep{Lattimer2012, Ozel2012} utilizing time delay analysis of pulses. To study the effect of anisotropy in Durgapal-Fuloria model under $f(Q)$ gravity framework { in background of both dS and AdS spacetime}, we have chosen some pulsars which are PSR J074 + 6620 \citep{r1-PSRJ074+6620} with low mass star as companion, PSR J1810+1744 \citep{r1-PSRJ1810+1744,r1-PSRJ1810+1744-1} with very low mass star as companion, PSR J1959+2048 (the black widow) \citep{star1} with PSR J2215+5135 (the redback) \citep{star1} as companion and GW190814 \citep{waves3} with a black hole as companion. This set of pulsars having masses greater than 2 $M_{\odot}$ are considered to be very massive. Inevitably the existence of such massive pulsars is supported by stiff EOS which allows comparatively larger values of pressure for different fixed values of energy density. In this scenario, the pressure dominates in the neutron star configuration and resists the gravitational collapse of the matter. 

\begin{figure}
\centering
\includegraphics[height=9.5cm,width=12.5cm]{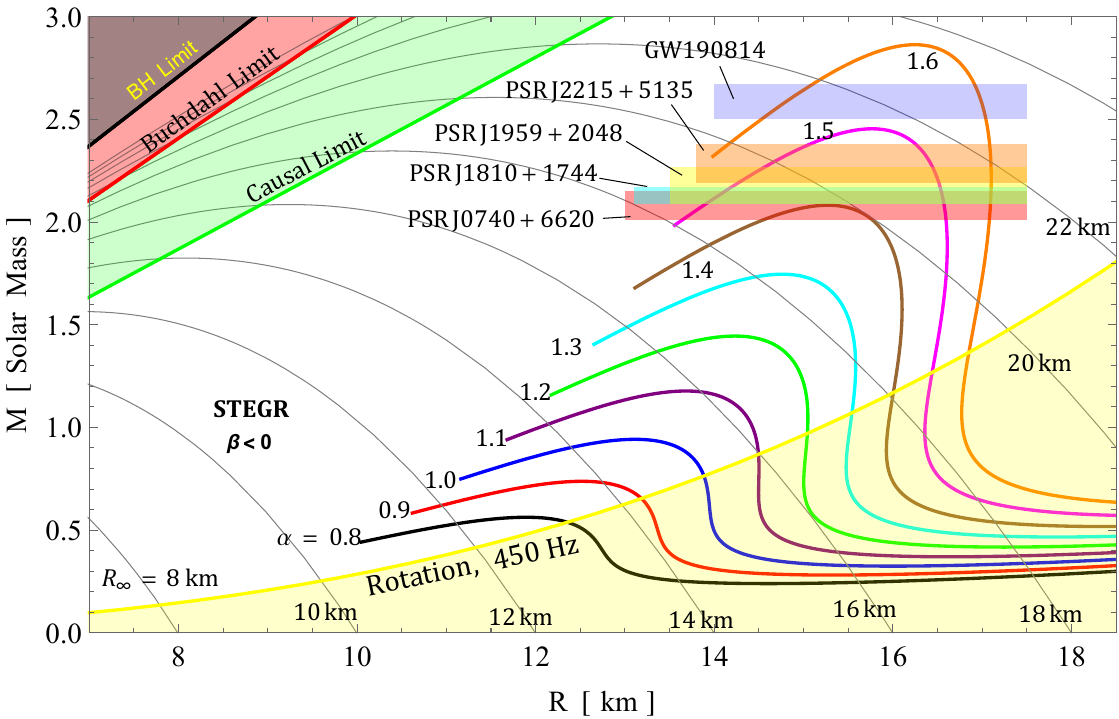}\\ \vspace{0.5cm}
\includegraphics[height=9.5cm,width=12.5cm]{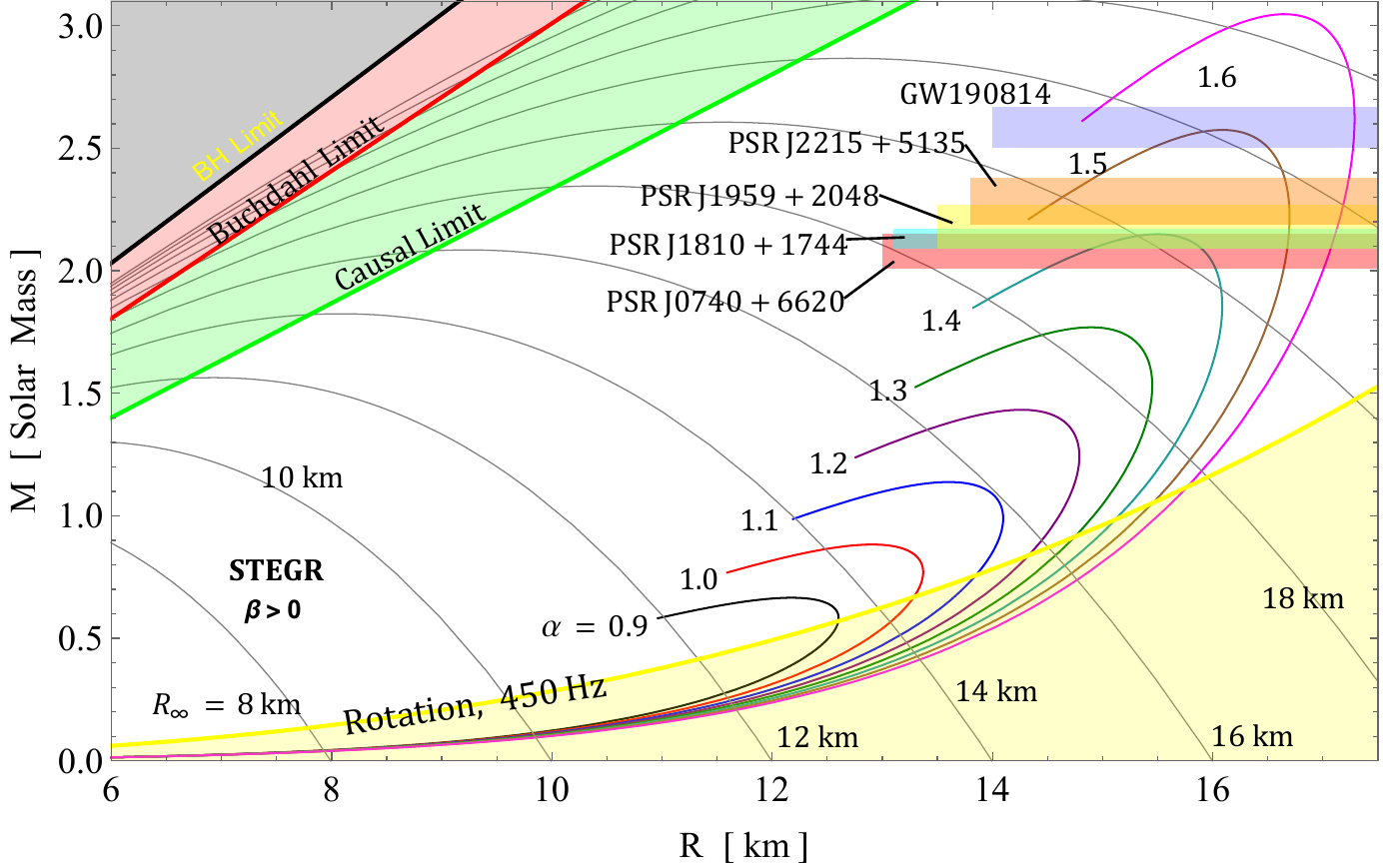}
\caption{$M-R$ curve for different values of $\alpha$ for $A= 0.0006~\text{km}^{-2}$, $\Delta_0 = 0.1$, $\beta = \pm 0.0004~\text{km}^{-2}$. Solid and lighter black curves represent radiation radius ($R_{\infty}$) contours. The light orange region represents the mass-shedding limit of 450 Hz pulsar.} 
\label{f4}
\end{figure}

\begin{figure}
\centering
\includegraphics[height=9.5cm,width=12.5cm]{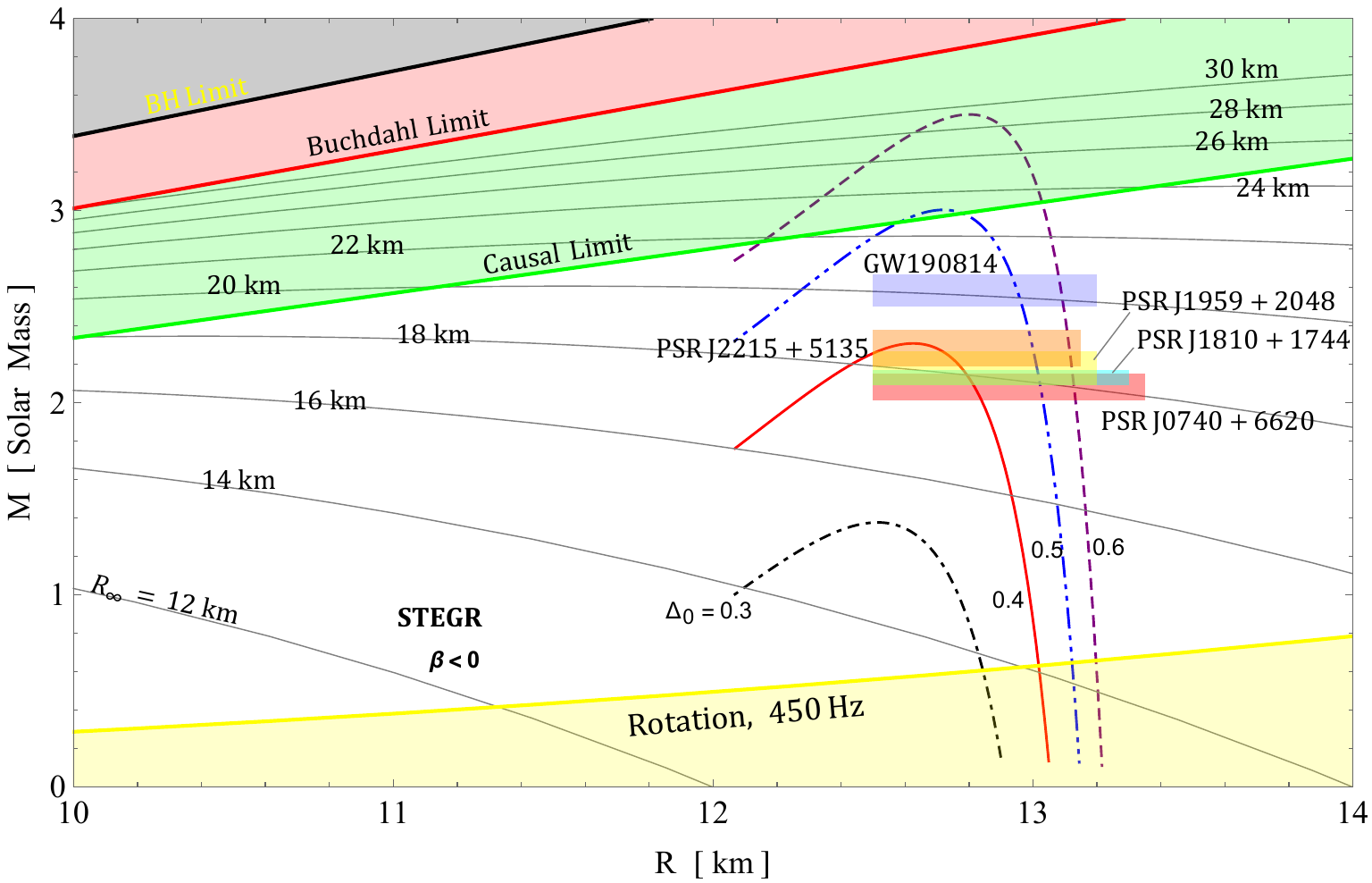}\\
\vspace{0.5cm}
\includegraphics[height=9.5cm,width=12.5cm]{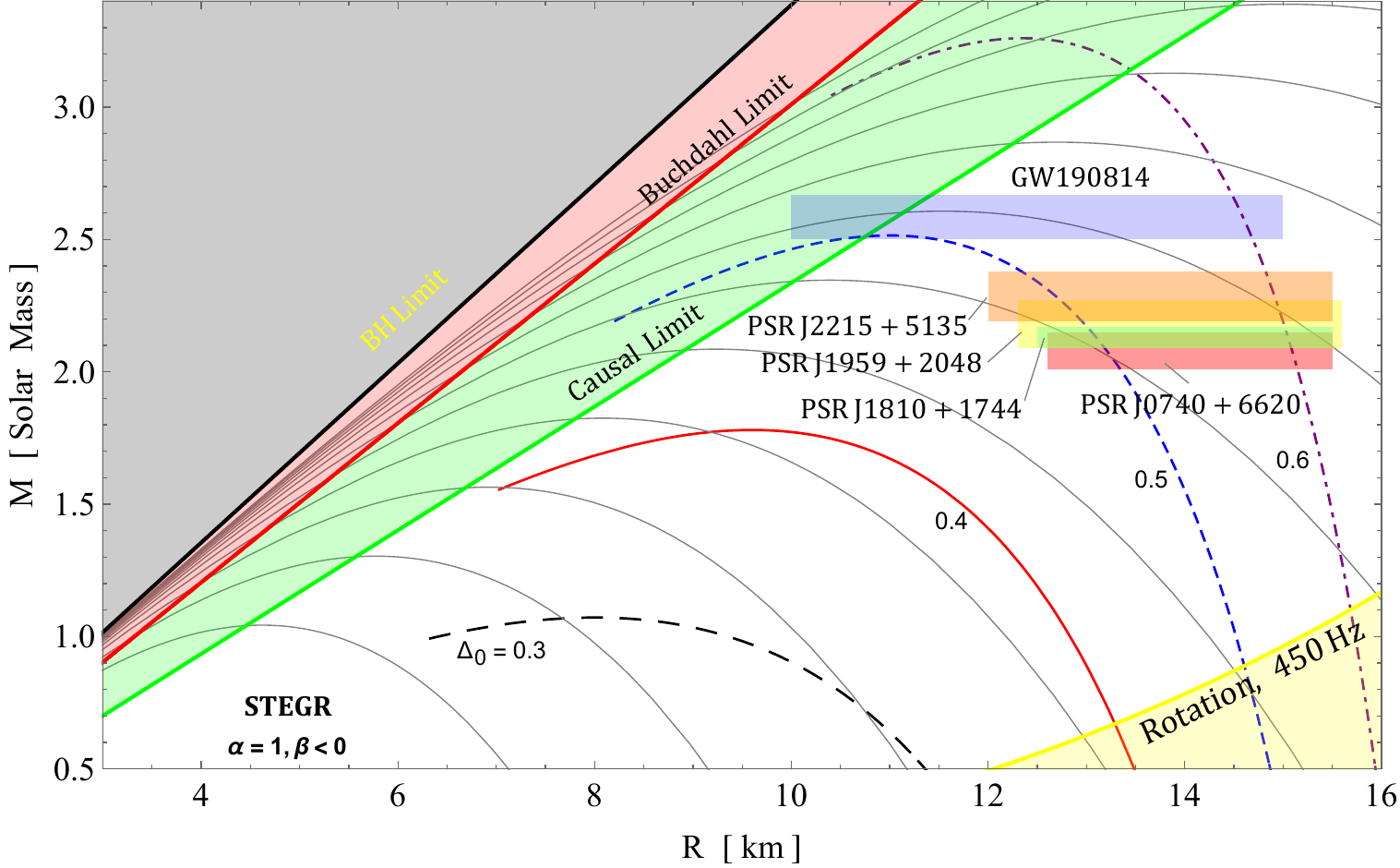}
\caption{$M-R$ curve for $\alpha=(0.5,1.0)$ for $A= 0.0006~\text{km}^{-2}$, $\beta = -0.0004~\text{km}^{-2}$ and different values of $\Delta_0$. Solid and lighter black curves represent radiation radius ($R_{\infty}$) contours. The light orange region represents the mass-shedding limit of 450 Hz pulsar.} 
\label{f4a}
\end{figure}

\begin{figure}
\centering
\includegraphics[height=9.5cm,width=12.5cm]{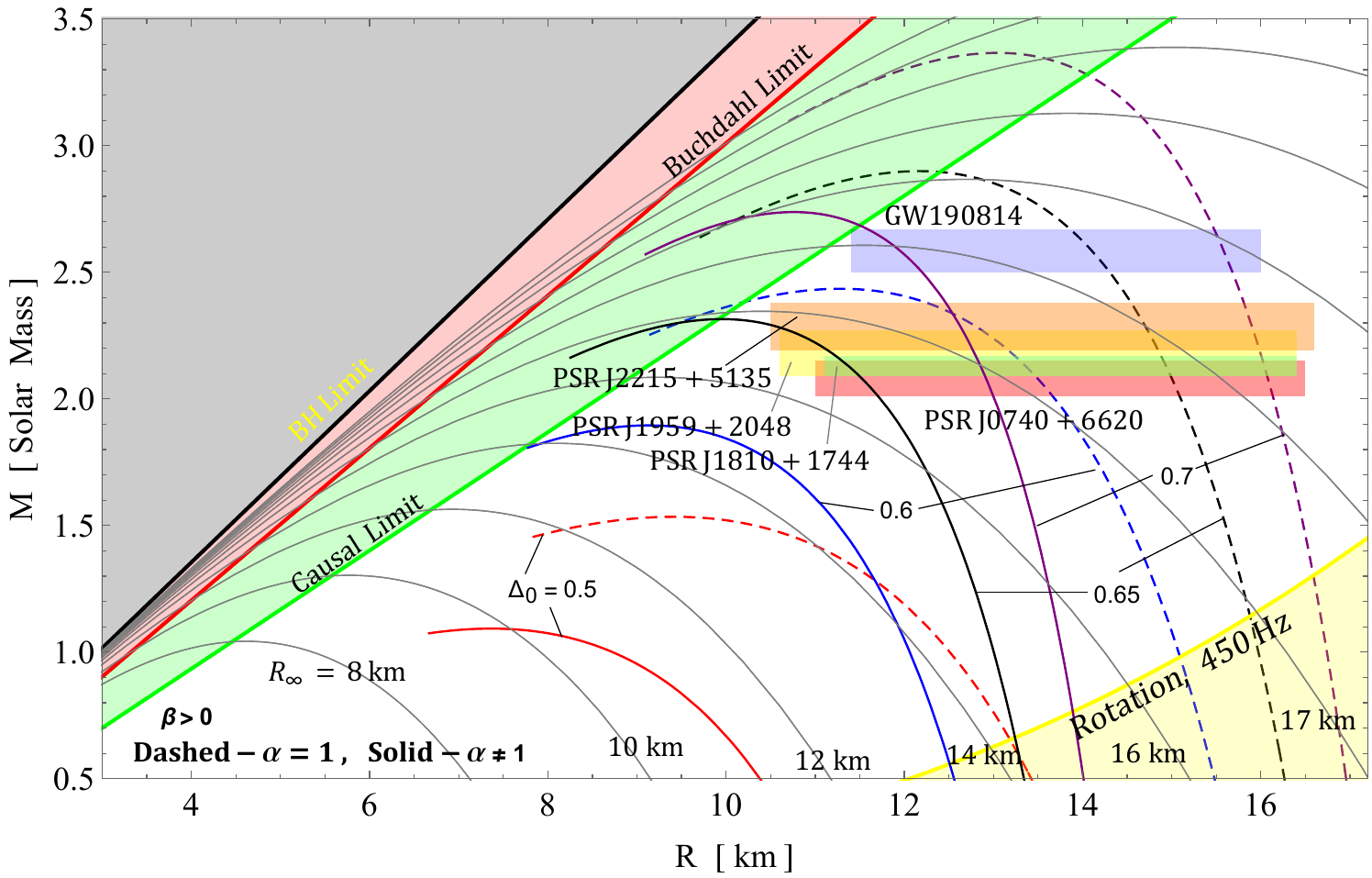}
\caption{$M-R$ curve for different values of $\Delta_0$ for $A= 0.0006~\text{km}^{-2}$, $\alpha = 0.5, 1.0$, $\beta = 0.0004~\text{km}^{-2}$. Solid and lighter black curves represent radiation radius ($R_{\infty}$) contours. The light orange region represents the mass-shedding limit of 450 Hz pulsar.} 
\label{f4b}
\end{figure}

\begin{figure}
\centering
\includegraphics[height=9.5cm,width=12.5cm]{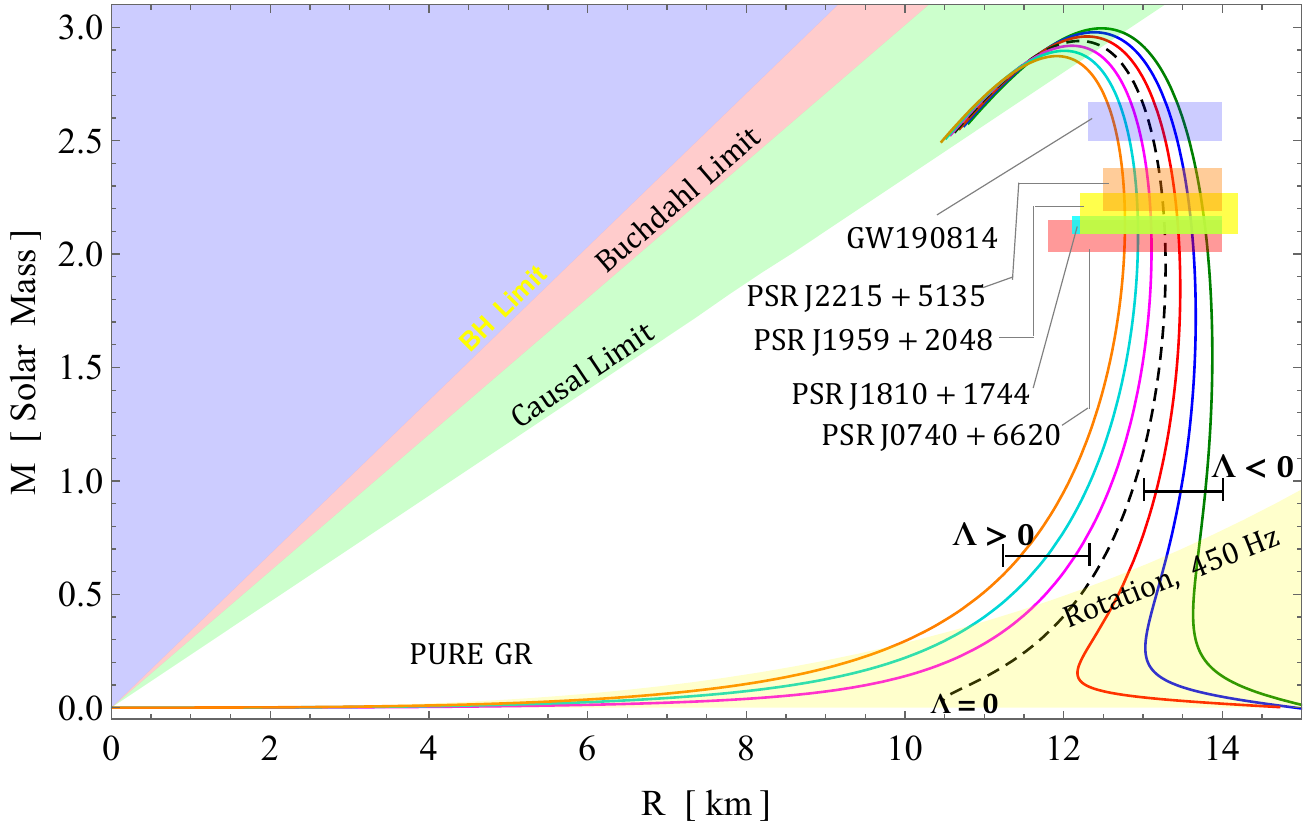}
\caption{$M-R$ curve for different values of $\Lambda$ with $A= 0.0028~\text{km}^{-2}$ in pure GR case. The light orange region represents the mass-shedding limit of 450 Hz pulsar.} 
\label{f4c}
\end{figure}

Measurements of radii of neutron stars seem more difficult and tricky \citep{Fortin2015,Ozel2016a} with observational parameters as compared to the determination of their masses. Certainly, both the tidal deformability factor related to the detection of gravitational waves \citep{Abbott2018} and NICER measurements on hotspots \citep{Miller2019, Riley2021} of neutron surfaces have opened ways to determine radii of neutron stars. For instance,  the radius of PSR J0740+6620 has been measured to be $13.70^{+2.6}_{-1.5}$ km  \citep{Miller2021} at 68\% credibility from NICER and XMM-Newton observations of rotating patterns of hotspots. Another reliable way is to determine the radiation radius which is given by $R_{\infty} = R \left(1- \frac{2M}{R}\right)^{-1/2}$ as a consequence of redshifted stars’ surface temperatures and luminosities occurring from the observed thermal emissions of neutron stars. For a physical star $R<R_{\infty}$.

 The determination of neutron star radii is a challenging task that can be eased by the progress in theoretical modeling. In this wor,k we have developed a framework to construct models of pulsars based on the well-behaved Durgapal-Fuloria ansatz in $f(Q)$ gravity. {Importantly, mass-radius curves are shown in Figure~\ref{f4} for different values of $\alpha$ for fixed anisotropy parameter (top panel $\beta<0$ and bottom panel $\beta>0$)}. It is to be noted that the $M-R$ curves correspond to the EOSs of matter via the TOV equation.  { The $M-R$ curves corresponding to  various values of $\alpha$ in Fig. \ref{f4} represent the STEGR case. For the chosen set of the constant parameters, the STEGR can hold very small mass for $\alpha\leq1$. In particular, for $\alpha = 1$ case we get $(0.939M_\odot,13.19km)$ for $\beta<0$ and ($0.88M_\odot,13.02km$) for $\beta>0$.} Consequently, one has to choose higher values of $f(Q)-$parameter $\alpha>1$ in order to fit and predict the radii of massive neutron stars.  It is to be noted that the pattern of $M-R$ curves changes when $\beta$ as well as $\Lambda$ changes from negative to positive value implying a change of EOS from polytropic types to quark matter equation of state. This signifies that the mass-radius relation and  EOS are influenced by positive $\Lambda$  and negative $\Lambda$ which is related to $f(Q)$-gravity parameters $\alpha$ and $\beta$. For fixed $\alpha$, depending on the sign, the effect of $\beta$ manifests itself in the background of dS and AdS spacetime with different $M-R$ relations and EOSs.  

\begin{table*}
\centering
\caption{The predicted radii of few high mass compact stars  corresponding Figure \ref{f4} in STEGR.}\label{table1}
\begin{minipage}{0.8\textheight}
 \scalebox{0.80}{\begin{tabular}{| *{8}{c|} }
\hline
 &    & \multicolumn{3}{c|}{{Predicted $R$ (km) ($\beta<0$)}} & \multicolumn{3}{c|}{{Predicted $R$ (km) ($\beta>0$)}}  \\[0.15cm]
\cline{3-8}
{Objects} & {$\frac{M}{M_\odot}$} & \multicolumn{3}{c|}{ $\alpha$ (STEGR)} & \multicolumn{3}{c|}{$\alpha$ (STEGR)} \\[0.15cm]
\cline{3-8}
&  & $1.4$ & $1.5$ & $1.6$ & $1.4$ & $1.5$ & $1.6$ \\[0.15cm] \hline
PSR J074 +6620 \citep{r1-PSRJ074+6620}  &  2.08 $\pm$ 0.07  & $15.39_{-0.10}^{+0.39}$  &  $16.57_{-0.02}^{+0.03}$  &   $17.10_{-0.01}^{+0.01}$  &  $15.91_{-0.36}^{+0.12}$  &  $16.67_{-0.03}^{+0.01}$  &   $17.06_{-0.06}^{+0.05}$  \\[0.15cm]
\hline
PSR J1810+1744 \citep{r1-PSRJ1810+1744-1} & 2.13$\pm$0.04  & -  &  $16.55_{-0.02}^{+0.03}$  &   $17.11_{-0.01}^{+0.01}$  &  $15.80_{-}^{+0.22}$  &  $16.68_{-0.04}^{+0.01}$  &   $17.10_{-0.04}^{+0.01}$ \\[0.15cm]
\hline
PSR J1959+2048 \citep{star1} & 2.18$\pm$0.09  & -  &  $16.52_{-0.07}^{+0.03}$  &   $17.10_{-0.01}^{+0.01}$  &  -  &  $16.69_{-0.02}^{+0.01}$  &   $17.11_{-0.03}^{+0.07}$ \\[0.15cm]
\hline
PSR J2215+5135 \citep{star1} & $2.28^{+0.10}_{-0.09}$ & -  &  $16.41_{-0.15}^{+0.11}$  &   $17.10_{-0.02}^{+0.01}$  &  -  &  $16.70_{-0.06}^{+0.01}$  &   $17.18_{-0.05}^{+0.06}$ \\[0.15cm]
\hline
GW190814 \citep{waves3} & 2.5-2.67 & -  &  -  &   $16.98_{-0.06}^{+0.06}$  &  -  &  $16.22_{-}^{+0.30}$  &   $17.28_{-0.01}^{+0.01}$  \\[0.15cm]
\hline
\end{tabular}}
 \end{minipage}
\end{table*}

\begin{table*}
\centering
\caption{The predicted radii of few high mass compact stars corresponding Figure \ref{f4a} in STEGR.}\label{table2}
\begin{minipage}{0.8\textheight}
 \scalebox{0.80}{\begin{tabular}{| *{7}{c|} }
\hline
 &    & \multicolumn{3}{c|}{{Predicted $R$ (km) ($\beta<0$)}} & \multicolumn{2}{c|}{{Predicted $R$ (km) ($\beta<0$)}}  \\[0.15cm]
\cline{3-7}
{Objects} & {$\frac{M}{M_\odot}$} & \multicolumn{3}{c|}{ $\Delta_0$ ($\alpha \neq 1$)} & \multicolumn{2}{c|}{$\Delta_0$ ($\alpha = 1$)} \\[0.15cm]
\cline{3-7}
&  & $0.4$ & $0.5$ & $0.6$ & 0.5 & 0.6 \\[0.15cm] \hline
PSR J074 +6620 \citep{r1-PSRJ074+6620}  &  2.08 $\pm$ 0.07 &  $12.82_{-0.03}^{+0.02}$ & $13.02_{-0.01}^{+0.01}$ & $13.13_{-0.01}^{+0.01}$ & $13.21_{-0.15}^{+0.12}$ & $15.09_{-0.04}^{+0.05}$  \\[0.15cm]
\hline
PSR J1810+1744 \citep{r1-PSRJ1810+1744-1} & 2.13$\pm$0.04 &  $12.80_{-0.02}^{+0.01}$ & $13.02_{-0.02}^{+0.01}$ & $13.12_{-0.01}^{+0.01}$ & $13.09_{-0.07}^{+0.10}$ & $15.05_{-0.04}^{+0.03}$ \\[0.15cm]
\hline
PSR J1959+2048 \citep{star1} & 2.18$\pm$0.09  &  $12.77_{-0.05}^{+0.04}$ & $13.01_{-0.01}^{+0.01}$ & $13.12_{-0.01}^{+0.01}$ & $12.97_{-0.22}^{+0.22}$ & $15.01_{-0.08}^{+0.08}$ \\[0.15cm]
\hline
PSR J2215+5135 \citep{star1} & $2.28^{+0.10}_{-0.09}$  &  - & $13.00_{-0.01}^{+0.01}$ & $13.11_{-0.01}^{+0.01}$ & $12.69_{-0.33}^{+0.26}$ & $14.91_{-0.08}^{+0.09}$ \\[0.15cm]
\hline
GW190814 \citep{waves3} & 2.5-2.67  &  - & $12.95_{-0.02}^{+0.02}$ & $13.09_{-0.01}^{+0.01}$  & $-$ & $14.60_{-0.11}^{+0.09}$ \\[0.15cm]
\hline
\end{tabular}}
 \end{minipage}
\end{table*}
\begin{table*}
\centering
\caption{The predicted radii of few high mass compact stars  corresponding Figure \ref{f4b} in STEGR.}\label{table3}
\begin{minipage}{0.8\textheight}
 \scalebox{0.80}{\begin{tabular}{| *{7}{c|} }
\hline
 &    & \multicolumn{2}{c|}{{Predicted $R$ (km) ($\beta>0$)}} & \multicolumn{3}{c|}{{Predicted $R$ (km) ($\beta>0$)}}  \\[0.15cm]
\cline{3-7}
{Objects} & {$\frac{M}{M_\odot}$} & \multicolumn{2}{c|}{ $\Delta_0$ ($\alpha \neq 1$)} & \multicolumn{3}{c|}{$\Delta_0$ ($\alpha = 1$)} \\[0.15cm]
\cline{3-7}
&  & $0.65$ & $0.70$ & 0.60 & 0.65 & 0.70 \\[0.15cm] \hline
PSR J074 +6620 \citep{r1-PSRJ074+6620}  &  2.08 $\pm$ 0.07 &  $11.53_{-0.22}^{+0.18}$ & $12.92_{-0.08}^{+0.07}$ & $13.49_{-0.16}^{+0.17}$ & $15.00_{-0.08}^{+0.09}$ & $16.08_{-0.05}^{+0.05}$  \\[0.15cm]
\hline
PSR J1810+1744 \citep{r1-PSRJ1810+1744-1} & 2.13$\pm$0.04 &  $11.36_{-0.12}^{+0.13}$ & $12.84_{-0.04}^{+0.06}$ & $13.36_{-0.11}^{+0.11}$ & $14.93_{-0.04}^{+0.06}$ & $16.04_{-0.03}^{+0.03}$ \\[0.15cm]
\hline
PSR J1959+2048 \citep{star1} & 2.18$\pm$0.09  &  $11.19_{-}^{+0.04}$ & $12.79_{-0.13}^{+0.11}$ & $13.21_{-0.67}^{+0.26}$ & $14.86_{-0.13}^{+0.13}$ & $15.99_{-0.06}^{+0.07}$ \\[0.15cm]
\hline
PSR J2215+5135 \citep{star1} & $2.28^{+0.10}_{-0.09}$  &  $10.62_{-}^{+0.46}$ & $12.64_{-0.17}^{+0.14}$ & $12.83_{-0.54}^{+0.36}$ & $14.72_{-0.16}^{+0.11}$ & $15.90_{-0.09}^{+0.08}$ \\[0.15cm]
\hline
GW190814 \citep{waves3} & 2.5-2.67  &  - & $12.01_{-0.41}^{+0.20}$ & $-$  & $14.14_{-0.53}^{+0.16}$ & $15.60_{-0.12}^{+0.09}$ \\[0.15cm]
\hline
\end{tabular}}
 \end{minipage}
\end{table*}

\begin{table*}
\centering
\caption{The predicted radii of few high mass compact stars  corresponding Figure \ref{f4c} in Pure GR.}\label{table4}
\begin{minipage}{0.8\textheight}
 \scalebox{0.80}{\begin{tabular}{| *{9}{c|} }
\hline
 &    & \multicolumn{7}{c|}{{Predicted $R$ (km)}}  \\[0.15cm]
\cline{3-9}
{Objects} & {$\frac{M}{M_\odot}$} & \multicolumn{7}{c|}{ $\Lambda \times 10^{-4} / km^2$}  \\[0.15cm]
\cline{3-9}
&  & $6$ & $4$ & 2 & 0 & $-2$ & $-4$ & $-6$ \\[0.15cm] \hline
PSR J074 +6620 \citep{r1-PSRJ074+6620}  &  2.08 $\pm$ 0.07 &  $12.77_{-0.01}^{+0.01}$ & $12.92_{-0.01}^{+0.01}$ & $13.11_{-0.01}^{+0.01}$ & $13.27_{-0.01}^{+0.02}$ & $13.45_{-0.01}^{+0.01}$ & $13.63_{-0.02}^{+0.01}$ & $13.80_{-0.02}^{+0.01}$ \\[0.15cm]
\hline
PSR J1810+1744 \citep{r1-PSRJ1810+1744-1} & 2.13$\pm$0.04 &  $12.77_{-0.01}^{+0.01}$ & $12.93_{-0.01}^{+0.01}$ & $13.09_{-0.01}^{+0.02}$ & $13.26_{-0.01}^{+0.02}$ & $13.44_{-0.01}^{+0.01}$ & $13.61_{-0.01}^{+0.01}$ & $13.78_{-0.01}^{+0.02}$ \\[0.15cm]
\hline
PSR J1959+2048 \citep{star1} & 2.18$\pm$0.09  &  $12.78_{-0.01}^{+0.01}$ & $12.93_{-0.01}^{+0.01}$ & $13.09_{-0.01}^{+0.02}$ & $13.26_{-0.02}^{+0.02}$ & $13.42_{-0.02}^{+0.03}$ & $13.60_{-0.04}^{+0.02}$ & $13.77_{-0.05}^{+0.02}$ \\[0.15cm]
\hline
PSR J2215+5135 \citep{star1} & $2.28^{+0.10}_{-0.09}$  &  $12.76_{-0.02}^{+0.01}$ & $12.92_{-0.03}^{+0.01}$ & $13.08_{-0.04}^{+0.01}$ & $13.23_{-0.03}^{+0.03}$ & $13.39_{-0.04}^{+0.04}$ & $13.56_{-0.05}^{+0.03}$ & $13.71_{-0.04}^{+0.05}$ \\[0.15cm]
\hline
GW190814 \citep{waves3} & 2.5-2.67  &  $12.63_{-0.08}^{+0.05}$ & $12.78_{-0.08}^{+0.05}$ & $12.93_{-0.09}^{+0.05}$  & $13.08_{-0.08}^{+0.05}$ & $13.23_{-0.09}^{+0.05}$ & $13.37_{-0.09}^{+0.06}$ & $13.52_{-0.09}^{+0.07}$ \\[0.15cm]
\hline
\end{tabular}}
 \end{minipage}
\end{table*}

In the top panel of Figure \ref{f4}, the $M-R$ curves increase gradually towards a maximum or peak and then decrease sharply to a minimum thereafter remaining constant for increasing radii. This trend highlights the variation in radius for a range of different masses of neutron stars. We observe that neutron stars with a small mass ($<1 ~M_{\odot}$) have larger radii for different values of $\alpha$. The intermediate part of $M-R$ curves shows the smaller change in the radii. As $\alpha$ increases the peak of $M-R$ curves shifts upwards and horizontally towards the right in both panels of Figure \ref{f4}. This means higher values $\alpha$ lead to neutron stars with larger mass as well as larger radii. Evidently, $M-R$ curves with larger values of $\alpha$ can be related to a stiffer EOS which supports the existence of massive neutron stars. 

\begin{table}
\centering
\caption{$M_{max}$ and $R$ corresponding to Figure \ref{f4} and \ref{f4a}.}\label{table7}
 \scalebox{0.85}{\begin{tabular}{| *{6}{c|} }
\hline
$\alpha$ & $M_{max}$ & $R$ & $\Delta_0$ & $M_{max}$ & $R$ \\[0.15cm]
& ($M_\odot$) & (km) & & ($M_\odot$) & (km)\\[0.15cm]
\hline
0.8 & 0.565  & 12.10 & 0.3 & 1.372 & 12.53 \\
0.9 & 0.736  & 12.72 & 0.4 & 2.303 & 12.64 \\
1.0 & 0.939  & 13.22 & 0.5 & 3.001 & 12.73 \\
1.1 & 1.174  & 13.80 & 0.6 & 3.505 & 12.80 \\
1.2 & 1.442  & 14.31 & - & - & - \\
1.3 &  1.742 & 14.83 &-  & - & - \\
1.4 & 2.083 & 15.30 & - & - & - \\
1.5 & 2.456 & 15.84 & - & - & - \\
1.6 & 2.862  & 16.30 & - & - & - \\[0.15cm]
\hline
\end{tabular}}
\end{table}

Furthermore, Fig. \ref{f4} also suggests that prediction of the radii of millisecond pulsars with a mass greater than 2 $M_{\odot}$ for a fixed value of anisotropy constant ($\Delta_0 = 0.1$) is possible for $\alpha \geq 1.4$.  However, both panels of Figure \ref{f4a} and \ref{f4b}, imply that radii of very massive pulsars can be predicted for lower values of $\alpha$ if anisotropy is increased to a sufficient amount. So, in $f(Q)$ gravity framework, there is a driving competition between $\alpha$ and anisotropy to establish an impact on the stiffness of the EOS of matter. Hence, anisotropy has a significant influence on $M-R$ curves in $f(Q)$ gravity. Contours of different fixed values of radiation radius are illustrated in Figure \ref{f4}, \ref{f4a} and \ref{f4b}. The horizontal bands corresponding to the chosen pulsars have been shown in Fig \ref{f4}, \ref{f4a} and \ref{f4b}.  The $M-R$ curves that cross constraints like causality limit and Buchdahl limit and those do not intersect with the horizontal bands of pulsars can be excluded for radii measurements. This helps to predict the radii tabulated in Table~\ref{table1}, \ref{table2} and \ref{table3}. 

{ To make a comparative analysis of the $M-R$ curves of STEGR with that of the GR, we have reassessed the present study by recomputing  all the expressions of density and pressures (see Appendix A) to obtain $M-R$ curves in the framework of pure general relativity with the presence of $\Lambda$.  In Fig. 7, one can see the variation of mass with radius for both positive and negative $\Lambda$ and the predicted radii corresponding to the stellar candidates are shown in Table 4.  We can see from Figs. 4-7  that the effect of $\beta$ (STEGR) or $\Lambda$ (Pure GR) on $M-R$ curves remain the same i.e. for $(\beta,\, \Lambda)>0$ has quark-like equation of state while $(\beta,\, \Lambda)<0$ has baryonic-like equation of state. One can also compare the STEGR and GR predictions of radii  from the Tables 1-4. Notably radii of the compact stars can be between $\approx 12.6-16\,km$ in STEGR with $\alpha=1$ while in Pure GR it can have radii between $\approx 12.6-13.8\,km$. Hence, pure GR has narrower radius range than the case $\alpha=1$ in STEGR.}

The intermediate linear part of $M-R$ curves goes to a certain height and sees a quick bend to go down with smaller changes in radii. The physical indication of the maxima of this bend is that matter within the neutron star is squeezed to its maximum limit. So, this peak provides the information on the maximum mass of the neutron star for a fixed value of $\alpha$ and anisotropy. The maximum masses and the corresponding radii for different values of $\alpha$ and $\Delta_0$ are in Table \ref{table7}. As compared to $M-R$ curves in  top panel of Figure \ref{f4}, similar kinds of $M-R$ curves with a maximum mass of 2.2 $M_\odot$ can be observed in earlier research work \citep{Togashi2016} based on variational methods and EOSs for nucleonic matter. Astashenok and Odintsov~\cite{Astashenok:2020cfv} conducted an investigation where they anticipated supermassive compact stars with masses ranging from $M \approx 2.2 - 2.3~M_{\odot}$ and radii of roughly $R_s \approx 11~\text{km}$ within the context of axion $R^2$ gravity. The presence of an axion ``galo'' around the star leads to a slower decay of the scalar curvature compared to vacuum $R^2$ gravity. In the study conducted by Nashed et al.~\cite{Nashed:2021gkp}, the authors explore the application of higher-order curvature theory to spherically symmetric spacetime solutions for the interior of stellar compact objects. The model successfully satisfies the required constraints for anisotropic compact stars, with particular emphasis on the Her X-1 star. The proposed model exhibits enhanced stability compared to analogous models in GR. Interestingly, the inclusion of hyperons in the EOS reduces the maximum mass effectively to 1.6 $M_\odot$. In other works \citep{Demorest2010, Lattimer2001}, different EOSs such as AP1-4 \citep{Akmal1997},  MS1-3 \citep{Muller1996}, PAL1-6 \citep{Prakash1988}, GM1-3 \citep{Glendenning1991} are utilized to plot $M-R$ curves which have similar type of patterns to $M-R$ curves in Figure \ref{f4} (top panel) produced in $f(Q)$ gravity.  It is worth mentioning that the EOSs such as MS2, PAL1, and AP3, AP4 are stiffer as these represent comparatively high mass neutron stars in the $M-R$ plot. 

\section{Stability Analysis} \label{sec7}

\subsection{Stability analysis via adiabatic index} \label{sec7.1}

Now, we must assess the stability of the anisotropic stellar configuration. To do this, it is necessary to examine the adiabatic index ($\Gamma$) given by
\begin{equation}
    \Gamma = \frac{\rho+P_r}{P_r}\frac{dP_r}{d\rho}.
\end{equation}

The stability criterion for an isotropic fluid in the Newtonian limit is expressed as $\Gamma > \frac{4}{3}$, as stated by Heintzmann and Bondi \citep{heintzmann1975neutron,bondi1992anisotropic} in their respective works on neutron stars and anisotropic fluids. The stability requirement for an anisotropic stellar model, deviates from the classic Chandrasekhar result for isotropic fluids~\citep{chan1992dynamical,chan1993dynamical},  
\begin{equation}
\Gamma > \frac{4}{3}\left(1 + \frac{\Delta}{r|(P_{r})|^{\prime}}+\frac{1}{4}\frac{\kappa \rho P_{r}r}{|(P_{r})|^{\prime}} \right),\label{eq62}
\end{equation}
where primes are used to represent differentiation with respect to the radial coordinate, denoted as $r$. The second component in Equation (\ref{eq62}) accounts for the adjustments in the stability condition caused by the presence of anisotropy ($P_r \neq P_t$), while the final term reflects relativistic adjustments. 

From Figure (\ref{f5}), it is evident that the non-metricity parameter $\alpha$ and the anisotropy parameter $\Delta_0$ significantly influence the limit for $\Gamma$ in the stability condition of an anisotropic star model. The left panel of Figure \ref{f5} illustrates that the value of $\Gamma$ increases as the value of $\alpha$ decreases. This indicates that the parameter $\alpha$ in the function $f(Q)$ determines the stability of the stellar structure. However, the value of $\alpha$ cannot be arbitrarily high, despite other physical parameters such as pressure, density, casual limit and compactness exhibiting well-behaved characteristics within the model. In the given scenario, the left panel of Figure \ref{f5} demonstrates that the value of $\Gamma$ is more than 1.33 when $\alpha$ is less than or equal to 13.8, which serves as the maximum limit for $\alpha$ in order to obtain a stable model. In addition, the right panel of Figure \ref{f5} exhibits the same pattern as the left panel, indicating that an increase in $\Delta_0$ corresponds to a decrease in the value of $\Gamma$. This suggests that a model with significant anisotropy may result in an unstable model. In this scenario, it may be inferred that a stable star configuration cannot hold an arbitrarily high level of anisotropy. Observing the bottom panel of Figure~\ref{f4}, it is evident that the mass-radius relationship of our model fails to satisfy the casualty condition when $\Delta_0\ge 0.5$. Based on the information shown in Figure~\ref{f4} and \ref{f5}, we can deduce that our model is stable when the value of $\Delta_0$ is less than 0.5.          
\begin{figure*}[!htb]
\centering
\includegraphics[height=6cm,width=7.5cm]{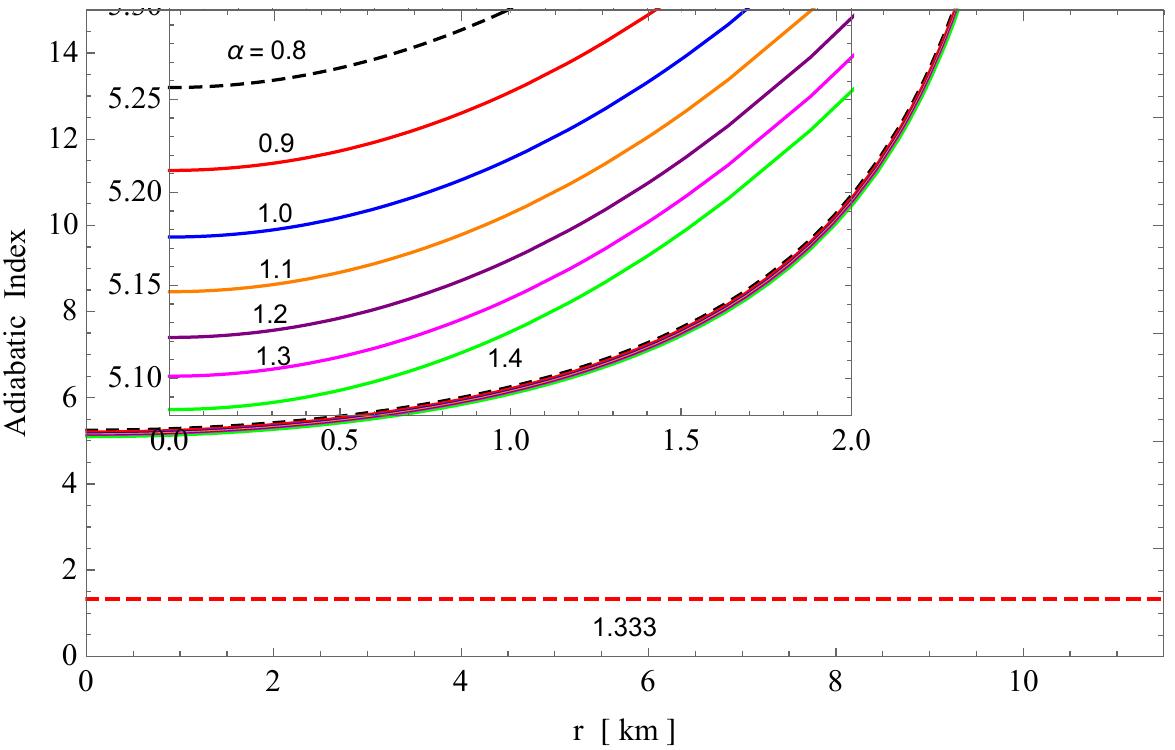}~~~~~~
\includegraphics[height=6
cm,width=7.5cm]{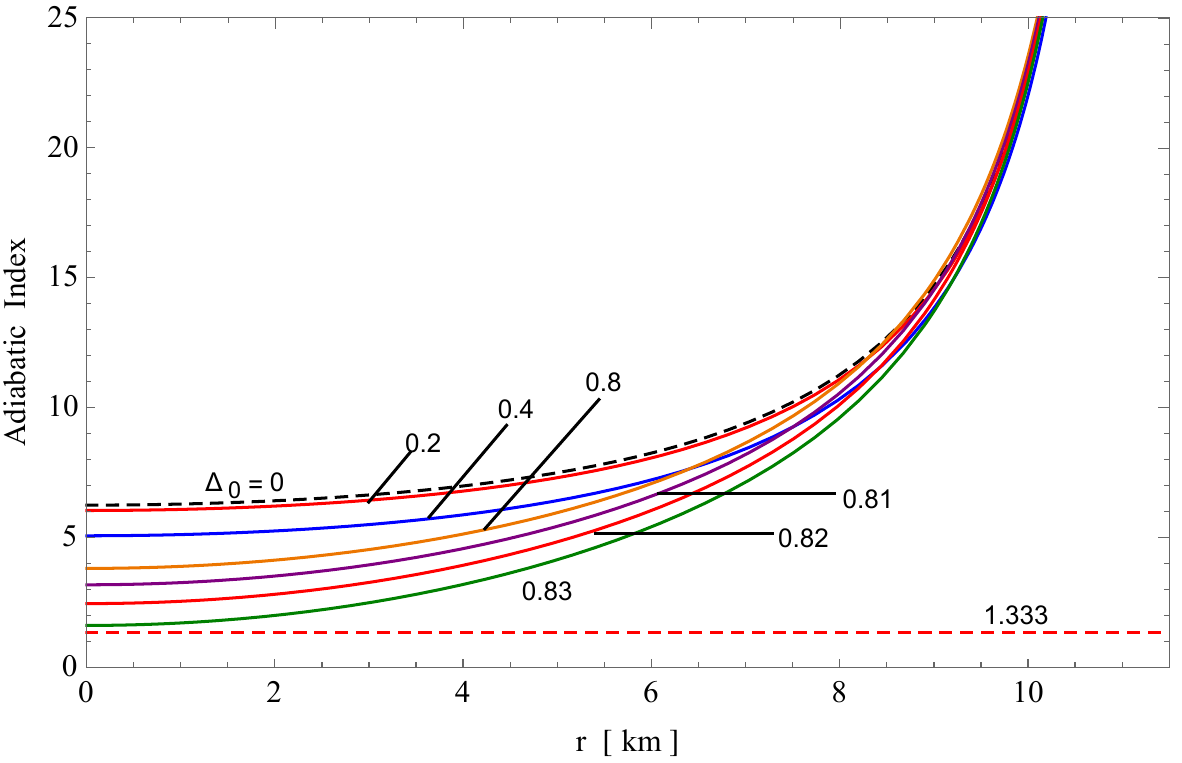}
\caption{Variation adiabatic index with respect to radius for different values of $\alpha$ for $A= 0.0006~\text{km}^{-2}$, $\Delta_0 = 0.35$, $\beta = -0.0004~\text{km}^{-2}$ and different values of $\Delta_0$ for $A= 0.0006~\text{km}^{-2}$, $\alpha = 0.5$, $\beta = -0.0004~\text{km}^{-2}$.}
\label{f5}
\end{figure*}

\subsection{Stability analysis using the Harrison-Zel'dovich-Novikov criterion} \label{sec7.2}

In a previous study, Chandrasekhar~\cite{chandrasekhar1964dynamical} presented a way to determine the stability of a star system when subjected to radial perturbations. The perturbed physical parameters in this scenario, such as the metric functions, pressure, and density, are provided as follows: 
\begin{eqnarray}
&& \hspace{-1.5cm} X \rightarrow X+\delta X, ~~~~Y \rightarrow Y+\delta Y\label{eq64}\\
&& \hspace{-1.5cm} P_r \rightarrow P_r+\delta P_r~,~\rho \rightarrow \rho+\delta \rho. \label{eq66}
\end{eqnarray}
The radial perturbations are modelled as exponential functions of the kind $e^{i\sigma t}$, where $\sigma$ represents the characteristic frequency and the perturbations exhibit oscillatory behaviour. The oscillation's stability is governed by the parameter $\sigma$. Therefore, we get from Eqs.~(\ref{eq17}) and (\ref{eq18}),  
\begin{small}
\begin{eqnarray}
\sigma^2 e^{Y-X} (P_r+\rho)\xi &=& {d(\delta P_r) \over dr}+\delta P_r {d \over dr} \left({Y \over 2}+X\right) +{\delta \rho \over 2} {dX \over dr}-{P_r+\rho \over 2}\left({dX \over dr}+{1 \over r}\right) \nonumber \\
&& \left({dY \over dr}+{dX \over dr}\right) \xi. ~\label{eq67}
\end{eqnarray}
\end{small}
The Lagrangian displacement, denoted by $\xi$, is directly related to the radial velocity $v$ by the equation $v=\partial \xi/\partial t$, where $t$ represents the world time.  

Moreover, the fluctuation in energy density may be mathematically represented as
\begin{equation}
\delta \rho = -{1 \over r^2} {\partial \over \partial r} \big[r^2(P_r+\rho)\xi \big] \label{eq68}
\end{equation}
stating the conservation of the baryon number as
\begin{eqnarray}
\delta P_r = -\xi {dP_r \over dr}-\Gamma_r P_r~ {e^{X/2} \over r^2} {d \over d r} \big[r^2 e^{-X/2} \xi \big], \label{eq69}
\end{eqnarray}
here, $\Gamma_r$ represents the radial adiabatic index. 

By replacing equations \eqref{eq68} and \eqref{eq69} in equation \eqref{eq67}, we find 
\begin{small}
\begin{eqnarray}
&& \hspace{-0.5cm} \sigma^2 e^{Y-X} (P_r+\rho)\xi = -{d \over dr} \left({\xi~{dP_r \over dr}} \right)-\left({1 \over 2} {dY \over dr}+{d X \over dr}\right) \xi ~{dP_r \over dr}-{P_r+\rho \over 2} \left({dX \over dr}+{1 \over r}\right)\left({dY \over dr}+{dX \over dr}\right) \xi \nonumber \\
&& \hspace{-0.5cm} -{1 \over 2} {dX \over dr} \Big\{{d \over dr} \Big[(P_r+\rho)\xi \Big]+{2(P_r+\rho)\xi \over r} \Big\}  -e^{-(Y+2X)/2} {d \over dr} \Big[e^{(Y+2X)/2}{\Gamma_r P_r \over r^2}~ e^{X/2} {d \over dr} \big(r^2 \,e^{-X/2}\xi \big)\Big], \label{eq70}
\end{eqnarray}   
\end{small}
which reduced to a more basic or straightforward form 
\begin{footnotesize}
\begin{eqnarray}
\sigma^2 e^{Y-X} (P_r+\rho)\xi &=& {4\xi \over r} {{dP_r \over dr}}-{e^{-{Y \over 2}} \over e^{X}} {d \over dr} \Big[e^{{(Y+3X)\over 2}} {\Gamma_r P_r \over r^2} {d \over dr} \big(r^2 \,e^{-X/2}\xi\big)\Big] +8\pi e^Y P_r(P_r+\rho) \xi-\nonumber \\
&& {\xi \over P_r+\rho} \left({dP_r \over dr}\right), \label{eq71}
\end{eqnarray}
\end{footnotesize}
This has been referred to the pulsation equation in the literature. 

Multiplying Eq. (\ref{eq71}) by $r^2 \xi e^{(Y+X)/2}$ yields 
\begin{footnotesize}
\begin{eqnarray}
&&  \sigma^2 \int_0^R e^{(3Y-X)/2} (P_r+\rho) r^2 \xi^2 dr = 4 \int_0^R e^{(Y+X)/2}  {dP_r \over dr} r \,\xi^2\,dr +  \int_0^R e^{(Y+3X)/2} {\Gamma_r P_r \over r^2} \left[{d \over dr}  \big(r^2 \,e^{-X/2}\xi \big)\right]^2dr \nonumber \\
&& \hspace{3cm} +\int_0^R e^{(3Y+X)/2} P_r (P_r+\rho) r^2 \xi^2 dr- \int_0^R e^{(Y+X)/2}{r^2 \xi^2 \over P_r+\rho} \left({dP_r \over dr}\right)^2 dr,\label{eq72}
\end{eqnarray}
\end{footnotesize}
as the ``{\it characteristic equation}''. This implies that the characteristic frequency $\sigma^2$ is positive for a small radial oscillation that does not collapse. Moreover, Chandrasekhar's computation is streamlined for the polytropic EOS, denoted as $P_r=K\epsilon^{\Gamma_r}$, as shown in references \citep{harrison1965gravitation} and ~\citep{zel1972relativistic}. It is proven that $\sigma^2>0$ for $\Gamma_r>4/3$. Therefore, the mass may be expressed as a mathematical function of its central density, given by $M(\rho_0) \propto \rho_0^{3(\Gamma_r-4/3)/2}$. Remarkably, we see that $M$ is a monotonically rising function of $\rho_0$ for $\Gamma_r>4/3$, resulting in $dM/d\rho_0 >0$. The {\it static stability criteria}, also known as the Harrison-Zel'dovich-Novikov criterion, asserts that for an anisotropic stellar object to remain stable, the total mass ($M$) must increase as the centre density ($\rho_0$) increases. 

\begin{figure}[!htb]
\centering
\includegraphics[height=7cm,width=8.7cm]{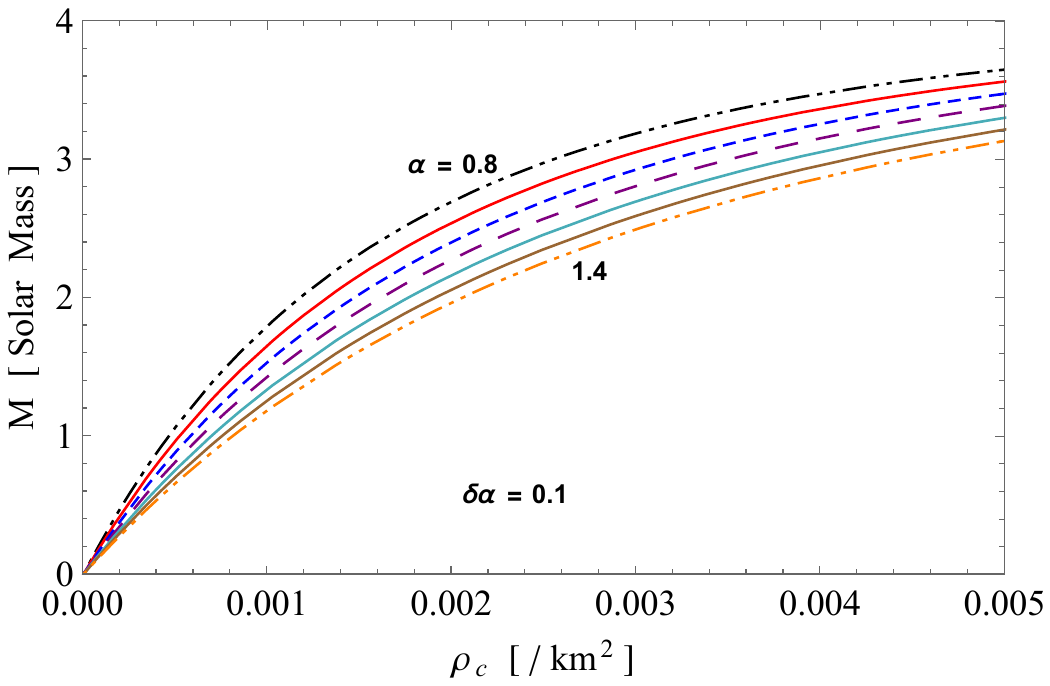}~~~~
\caption{Variation of mass with central density for different values of $\alpha$ for $A= 0.0006~\text{km}^{-2}$, $\Delta_0 = 0.1$, $\beta = -0.0004~\text{km}^{-2}$.}
\label{f6}
\end{figure}

In order to examine the condition for stability, we possess the following mathematical representations for the physical variables $M,~ \rho_0,~ dM/d\rho_0$, 
\begin{eqnarray}
\rho_0 &=& \frac{72 A \alpha}{7}-\frac{\beta}{2},  \label{eq73}\\
M &=&  \frac{1}{6 \left(144 \alpha +14 \rho_0 R^2+7 \beta  R^2\right)^2} \Big[R^3 \big(28 \rho_0^2 R^2 (24 -7 \Lambda  R^2)  \alpha  \rho_0 \left(36-7 \Lambda R^2\right)+28 \beta  \rho_0 R^2 \nonumber\\
&& (24-7 \Lambda R^2)-\Lambda \left(144 \alpha +7 \beta  R^2\right)^2+24 \beta  \left(432 \alpha +7 \beta  R^2\right)\big)\Big],\\
\frac{dM}{d\rho_0} &=& \frac{1152 \alpha  R^3 \left(432 \alpha -14 \rho_0 R^2-7 \beta  R^2\right)}{\left(144 \alpha +14 \rho_0 R^2+7 \beta  R^2\right)^3}
\end{eqnarray}

From Figure \ref{f6}, it is evident that mass is a monotonically increasing function of central density. Once again, Figure \ref{f6} demonstrates that the rate of change of mass in relation to the central density is positive and follows a linear pattern over the whole stellar area. Hence, the current model of anisotropic star satisfies the stability condition.

\section{Concluding Remarks} \label{sec8}
In this paper, we have explored the impact of anisotropy on the mass, radius, and stability of pulsars within the framework of $f(Q)$-gravity. By adopting the Durgapal-Fuloria ansatz described in (\ref{eq28}), we derived a viable expression for the anisotropy function, denoted as $\Delta$. This expression  allowed us to obtain a closed-form solution, as presented in (\ref{eq30}). Our findings demonstrated that several compact objects, such as PSR J074+6620 \citep{r1-PSRJ074+6620}, PSR J1810+1744 \citep{r1-PSRJ1810+1744, r1-PSRJ1810+1744-1}, PSR J1959+2048 \citep{star1}, PSR J2215+5135 \citep{star1}, and GW190814 \citep{waves3}, are in good agreement with observational data. These pulsars serve as specific examples that highlight the significance of anisotropy in shaping their properties. Next, by utilizing the chosen metric functions, we proceeded to simplify the Einstein field equations and investigated the macroscopic characteristics of anisotropic pulsars within the framework of the theory under consideration. To account for the local anisotropy in pressure within the pulsars, we incorporated the Bower \& Liang approach proposed in 1974 \citep{BL1974}. While there exists alternative formalisms such as the Horvat model \citep{Horvat:2010xf} and the Cosenza model \citep{Cosenza:1981myi} that can also incorporate anisotropy within pulsars, the Bower-Liang model stands out for its simplicity. It assumes that the anisotropy is gravitationally induced and exhibits a nonlinear variation with radial pressure.

By considering the physical constraints, we have successfully aligned the interior solution with an exterior Schwarzschild  de Sitter spacetime and Schwarzschild Anti-de Sitter spacetime while fixing the constants $C$ and $D$. Consequently, with the known values of these constant parameters, we further ascertained the masses and radii of pulsars. The models have been carefully analyzed to ensure they meet the necessary conditions for viability within the stellar fluid's interior. Notably, the energy density within the configuration was regular at every point and decreases monotonically as the radial coordinate increases, reaching its maximum at the center of the self-gravitating object. Increasing the magnitude of the constant, $\alpha$ led to a higher density in the core, effectively squeezing more matter into central, concentric shells. However, beyond the central region, changes in $\alpha$ did not significantly affect the stellar density. The radial pressure as a function of the radial coordinate was examined in detail. We investigated two scenarios: one with $\alpha = 0.5$ and another with $\alpha = 1$, while keeping the anisotropy parameter $\Delta_0$ constant. Additionally, we explored the effects of varying the anisotropy parameter by considering $\Delta_0 = 0.1$ and $\Delta_0 = 0.35$ while keeping $\alpha$ fixed. Similar to the density trend, we found that the radial pressure remains continuous throughout the star and decreased monotonically as the radial coordinate increased. Notably, when $\Delta_0 = 0.1$, the central region exhibited higher radial pressure compared to the case of $\Delta_0 = 0.35$ by a factor of $10^2$. Increasing the anisotropic factor appeared to result in a relaxation of the fluid particles, particularly in the central regions of the star. We delved into the characteristics of the anisotropy parameter, $\Delta(r) = P_r - P_t$, where a positive $\Delta(r)$ indicates a repulsive force dominated by radial stresses that counteracts the inward gravitational force, thereby contributing to the stability of the compact object. By varying $\alpha$ while keeping $\Delta_0$ constant, we found that the anisotropy behaved smoothly, vanishing at the center of the configuration and increasing in magnitude towards the outer boundary, resulting in enhanced stability in the surface layers of the star. Moreover, doubling $\alpha$ from $0.5$ to $1.0$ led to a doubling of anisotropy within the stellar body. When examining the variation of $\Delta(r)$ with changes in $\Delta_0$ while $\alpha$ remains fixed, we observed a tenfold increase in anisotropy at each interior point of the gravitating body as $\Delta_0$ increases. The parameter $\Delta_0$ plays a crucial role in rendering the stellar fluid more anisotropic, introducing a greater disparity between radial and tangential stresses, which ultimately aids in stabilizing the star. 

Next, the Durgapal-Fuloria model in $f(Q)$ gravity had been developed to study the effect of anisotropy on pulsars. Specific pulsars, including PSR J074+6620, PSR J1810+1744, PSR J1959+2048, PSR J2215+5135, and GW190814, had been considered for their high masses. These massive pulsars are supported by EOSs that allow for larger pressure values at fixed energy density, preventing gravitational collapse. To analyze the mass-radius relationship, $M-R$ curves have been constructed for different values of $\alpha$ (a parameter in the model) and anisotropy. The curves exhibited several notable features: the presence of contours representing fixed values of radiation radius, two bending parts at the top and bottom of the curves (with an almost linear part in the middle), and the dependence on both $\alpha$ and anisotropy. The curves showed that neutron stars with smaller masses have larger radii, and increasing $\alpha$ leads to neutron stars with larger masses and radii. The prediction of radii for pulsars with masses exceeding 2 $M_{\odot}$ is possible for certain values of $\alpha$ and anisotropy. The maximum mass of a neutron star can also be inferred from the peak of the bending part in the $M-R$ curves. Comparisons with other studies using several methods and different EOSs have shown similar patterns in the $M-R$ curves. For instance, earlier research work by Togashi et al. \citep{Togashi2016}, based on variational methods and EOSs for nucleonic matter, also exhibited $M-R$ curves with a maximum mass of 2.2 $M_\odot$. In a study by Astashenok and Odintsov \cite{Astashenok:2020cfv}, supermassive compact stars with masses around $2.2-2.3M_{\odot}$ and radii approximately $11\text{km}$ were predicted within the framework of axion $R^2$ gravity. The presence of an axion ``halo'' around the star resulted in a slower decay of the scalar curvature compared to vacuum $R^2$ gravity. In contrast, Nashed et al.~\cite{Nashed:2021gkp} conducted a study examining for spherically symmetric anisotropic stellar models within the framework of higher-order curvature theory, focusing specifically on the Her X-1 star. Their proposed model demonstrated enhanced stability compared to analogous models in GR. Interestingly, the inclusion of hyperons in the EOS effectively reduced the maximum mass to 1.6 $M_\odot$. Other works, such as those by authors \citep{Demorest2010,Lattimer2001,Akmal1997,Muller1996,Prakash1988,Glendenning1991}, and the EOSs they utilized (e.g., AP1-4, MS1-3, PAL1-6, GM1-3) exhibited similar patterns in the $M-R$ curves to those observed in $f(Q)$ gravity (as shown in Figure \ref{f4}). Notably, EOSs like MS2, PAL1, AP3, and AP4, representing higher-mass neutron stars, resulted in stiffer $M-R$ curves in the plot. The $M-R$ curves obtained in the Durgapal-Fuloria model in $f(Q)$ gravity resembled those produced by other approaches, indicating the stiffness of the EOS. 

{ We have made a comparative evaluation of the $M-R$ curves obtained both in STEGR and GR background. In Fig. 7, the dependence of mass-radius relationship on positive and negative $\Lambda$ have been shown and the predicted radii corresponding to the stellar candidates are shown in Table 4. Comparing Figs 4-7  that the effect of $\beta$ (STEGR) or $\Lambda$ (Pure GR) on $M-R$ curves remain the same i.e. for $(\beta,\, \Lambda)>0$ has quark-like equation of state while $(\beta,\, \Lambda)<0$ has baryonic-like equation of state. Tables 1-4 refer that radii of the compact stars can be between $\approx 12.6-16\,km$ in STEGR with $\alpha=1$ while in Pure GR it can have radii between $\approx 12.6-13.8\,km$. Hence, stars have higher compactness in pure GR  than the case of $\alpha=1$ in STEGR. Remarkably we have explored that two different description of gravity, i.e., GR (curvature based) and STEGR (non-metricity based) provide similar structures of supermassive observed stars with analogous effect of cosmological constant on $M-R$ relationship of compact stars.}

The stability of an anisotropic stellar configuration is determined by the adiabatic index ($\Gamma$), which must be greater than $4/3$. The stability criterion is influenced by anisotropy and relativistic adjustments. The constant $\alpha$ and $\Delta_0$ play a key role in the stability condition. Decreasing $\alpha$ increases $\Gamma$ and promotes stability, but $\alpha$ cannot be too high. For stability, $\alpha$ must be less than or equal to $13.8$, ensuring $\Gamma$ remains above $1.33$. Increasing $\Delta_0$ decreases $\Gamma$, indicating that excessive anisotropy leads to instability. A stable configuration cannot have $\Delta_0$ greater than $0.5$. The mass-radius relationship violates the casualty condition when  $\Delta_0\geq 0.5$. Therefore, a stable model requires $\Delta_0$ to be less than $0.5$. The stability of the oscillations was also determined by the characteristic frequency, with positive values indicating stability. The computation focused on the polytropic EOS and showed that stability is achieved when the polytropic index is greater than $4/3$. The mass of the star was found to be a monotonically increasing function of the central density, satisfying the static stability criteria. Mathematical representations of the mass, central density, and the rate of change of mass with respect to density were provided. The figures presented confirmed that the model satisfied the stability condition by showing the increasing mass and positive derivative of mass with respect to density.

Our findings have offered valuable insights into the intricate interplay between anisotropy and the gravitational interaction of $f(Q)$-gravity, advancing our comprehension of pulsars and their underlying physical mechanisms.\\

\textbf{Acknowledgments}: 
SKM is thankful for continuous support and encouragement from the administration of University of Nizwa. AE thanks the National Research Foundation of South Africa for the award of a postdoctoral fellowship.
G. Mustafa is very thankful to Prof. Gao Xianlong from the Department of Physics, Zhejiang Normal University, for his kind support and help during this research. Further, G. Mustafa acknowledges grant No. ZC304022919 to support his Postdoctoral Fellowship at Zhejiang Normal University.\\

\section*{Conflict Of Interest statement }
The authors declare that they have no known competing financial interests or personal relationships that could have appeared to influence the work reported in this paper.

\section*{Data Availability Statement} 
This manuscript has no associated data, or the data will not be deposited. (There is no observational data related to this article. The necessary calculations and graphic discussion can be made available
on request.)

{ \section*{A. Exact anisotropic solution in pure Einstein's general relativity (GR)} \label{appen}
Let us consider the Einstein field equation for anisotropic matter distribution in the presence of cosmological constant $\Lambda$ as 
\begin{eqnarray}
\mathcal{R}_{\epsilon \nu}-\frac{1}{2} g_{\epsilon \nu} \mathcal{R}+g_{\epsilon \nu} \Lambda= -8\pi T^G_{\epsilon \nu} ,
\end{eqnarray}
where $G=c=1$, and matter tensor is described as: $\Big[T^{\epsilon}_\nu\Big]^G=\text{diag}\{ \rho^G, -P^G_r, P^G_t, P^G_t\}$. 
\begin{eqnarray}
\label{eq7aa}
&&\hspace{-0.6cm}  \rho^G=\frac{1}{8\pi}\bigg[\frac{1}{r^2}-e^{-X}\Big(\frac{1}{r^2}-\frac{X^{\prime}}{r}\Big)\bigg]-\Lambda,
\\ 
\label{eq8aa}
&&\hspace{-0.6cm}  P^G_{r}=\frac{1}{8\pi}\bigg[-\frac{1}{r^2}+e^{-X}\Big(\frac{1}{r^2}+\frac{Y^{\prime}}{r}\Big)\bigg]+\Lambda,
\\
\label{eq9aa}
&&\hspace{-0.6cm}  P^G_t=\frac{1}{8\pi}\bigg[\frac{e^{-X}}{4}\Big(2Y^{\prime\prime}+Y^{\prime2}-X^{\prime}Y^{\prime}+\frac{2(Y^{\prime}-X^{\prime})}{r}\Big)\bigg]+\Lambda,~~~
\end{eqnarray} 
and corresponding pressure anisotropy equation gives,
\begin{eqnarray}
P^G_t-P^G_r=\frac{1}{8\pi}\bigg[\frac{e^{-X}}{4}\Big(2Y^{\prime\prime}+Y^{\prime2}-X^{\prime}Y^{\prime} -\frac{2X^{\prime}}{r}\Big)+\frac{1+e^{-X}(r Y^{\prime}-1)}{r^2} \bigg]. \label{eq4.5a}
\end{eqnarray}
Now using the same ansatz (\ref{eq28}) for $X(r)$, we find the solution for $Y(r)$ using a suitable form of $\Delta^G=(P^G_t-P^G_r)$. The procedure for solution of Einstein's system of Eqs. (\ref{eq7aa})-(\ref{eq4.5a}) and corresponding matter variables in context of pure GR given below:\\

Let us consider the expression for anisotropy $\Delta^G(r)$, 
\begin{eqnarray}\label{eq30a}
\Delta^G(r)={\frac {8 \Delta_0\,A^2r^{2}\big[(2+5\Delta_0)+(\Delta_0-16)A r^2-2A^2r^4\big]}{56 \pi  \left( 1+Ar^2
 \right) ^{3} \left( 1-\Delta_0+Ar^2 \right) ^{2}}}, 
\end{eqnarray}
and using (\ref{eq28}), we find the solution for $Y(r)$ by solving of Eq.~(\ref{eq4.5a}) as, 
\begin{eqnarray}
&& \hspace{-0.6cm}Y(r)= (1-\Delta_0+Ar^2)^4\Bigg[D_1\Bigg\{\frac{[{\mathcal{A}_1}+{\mathcal{A}_2}(1-\Delta_0+Ar^2)+{\mathcal{A}_3}(1-\Delta_0+Ar^2)^2]Y_1(r)}{(1-\Delta_0+Ar^2)^3} \nonumber\\ 
&& \hspace{-0.6cm}+\frac{1536-384\Delta_0+48\Delta_0^{2}-2\Delta_0^{3}}{3(16-8\Delta_0-\Delta_0^{2})^{7/2}}\
\ln\left[\frac{\mathcal{A}_1-(4+\Delta_0)(1-\Delta_0+Ar^{2})+
\sqrt{{\mathcal{A}_4}}Y_1(r)}{(1-\Delta_0+Ar^2){\Delta_{0 5}}}\right]\Bigg\}+D_2\Bigg]^2,~~~~~~~\label{eq3.8a}
\end{eqnarray}
where $D_1$ and $D_2$ are the arbitrary constants of integration and expressions for other coefficients are,
\begin{eqnarray}
&& \hspace{-0.5cm}  Y_1(r)=\sqrt{({\Delta_{0 5}}-2(4+\Delta_0)(1-\Delta_0+Ar^2)
-(1-\Delta_0+Ar^2)^2)}, \nonumber\\
&& \hspace{-0.5cm} 
{\mathcal{A}_1}=\frac{\Delta_0}{3(16-8\Delta_0-\Delta_0^{2})},~~~~
{\mathcal{A}_2}=\frac{24-2\Delta_0+\Delta_0^{2}}{3(16-8\Delta_0-\Delta_0^{2})^{2}},~~~~
{\mathcal{A}_3}=\frac{288+80\Delta_0-10\Delta_0^{2}+\Delta_0^{3}}{3(16-8\Delta_0-\Delta_0^{2})^{3}},\nonumber\\
&& \hspace{-0.5cm} \mathcal{A}_4=(16-8\Delta_0-\Delta_0^{2}). \nonumber
\end{eqnarray}
Then we may find the expressions for matter variables $\rho^G$, $P^G_r$, and $P^G_t$ in GR framework as,
\begin{eqnarray}
&&\rho^G=\frac{8A(9+2Ar^{2}+A^2r^{4})}{56 \pi (1+Ar^{2})^{3}} -\Lambda, \\
&& P^G_r= \frac{1}{56 \pi r \Psi(r) \left(A r^2+1\right)^2}\Big\{2 \alpha  \big[Y_{11}(r) \left(7-A r^2 \left(A r^2+10\right)\right)-4 A r Y(r) \left(A r^2+3\right)\big]\Big\}+\Lambda,~~~~~~~~\\
&& P^G_t=\frac{1}{56 \pi r Y(r) \left(A r^2+1\right)^2}\Big\{2 \alpha  \big[Y_{11}(r) \left(7-A r^2 \left(A r^2+10\right)\right)-4 A r Y(r) \left(A r^2+3\right)\big]\Big\}\nonumber\\&&\hspace{0.9cm}+{\frac {8 \Delta_0\,A^2r^{2}\big[(2+5\Delta_0)+(\Delta_0-16)A r^2-2A^2r^4\big]}{56 \pi  \left( 1+Ar^2
 \right) ^{3} \left( 1-\Delta_0+Ar^2 \right) ^{2}}}+\Lambda
\end{eqnarray}
where, $Y_{11}(r)=dY/dr$. Moreover, we can find the constant involve in solution using the boundary conditions.}

\end{document}